\documentclass[twocolumn,prb,superscriptaddress,longbibliography,amsmath,amssymb,]{revtex4-1}
\usepackage{bm,upgreek}
\usepackage{graphicx}
\usepackage{color} 
\usepackage{setspace}
\usepackage{dcolumn}
\usepackage{subfigure}
\usepackage{hyperref}
\newcounter{firstbib}

\begin{document}
\title{Guiding light via geometric phases}
\author{Sergei Slussarenko}
\thanks{These authors contributed equally to this work.}
\affiliation{Dipartimento di Fisica, Universit\`a di Napoli Federico II, Complesso Universitario di Monte S. Angelo, via Cintia, 80126 Napoli, Italy}
\affiliation{Centre for Quantum Dynamics and Centre for Quantum Computation and Communication Technology, Griffith University, Brisbane, Queensland 4111, Australia}
\author{Alessandro Alberucci}
\thanks{These authors contributed equally to this work.}
\affiliation{Nonlinear Optics and OptoElectronics Lab, University Roma Tre, I-00146 Rome, Italy}
\affiliation{Optics Lab, Department of Physics, Tampere University of Technology, FI-33101 Tampere, Finland}
\author{Chandroth P. Jisha}
\affiliation{Centro de F\'{\i}sica do Porto, Faculdade de Ci\^encias, Universidade do Porto, 4169-007 Porto, Portugal}
\author{Bruno Piccirillo}
\affiliation{Dipartimento di Fisica, Universit\`a di Napoli Federico II, Complesso Universitario di Monte S. Angelo, via Cintia, 80126 Napoli, Italy}
\author{Enrico Santamato}
\affiliation{Dipartimento di Fisica, Universit\`a di Napoli Federico II, Complesso Universitario di Monte S. Angelo, via Cintia, 80126 Napoli, Italy}
\author{Gaetano Assanto}
\affiliation{Nonlinear Optics and OptoElectronics Lab, University Roma Tre, I-00146 Rome, Italy}
\affiliation{Optics Lab, Department of Physics, Tampere University of Technology, FI-33101 Tampere, Finland}
\affiliation{Consiglio Nazionale delle Ricerche, Institute for Complex Systems (ISC), Via dei Taurini 19, 00185 Rome, Italy}
\author{Lorenzo Marrucci}
\affiliation{Dipartimento di Fisica, Universit\`a di Napoli Federico II, Complesso Universitario di Monte S. Angelo, via Cintia, 80126 Napoli, Italy}
\affiliation{Consiglio Nazionale delle Ricerche, Institute of Applied Science \& Intelligent Systems (ISASI), Via Campi Flegrei 34, 80078 Pozzuoli (NA), Italy}
\date{01/08/2016}

\begin{abstract}
\textbf{All known methods for transverse confinement and guidance of light rely on modification of refractive index, that is, on the scalar properties of electromagnetic radiation \cite{book_snyder_1983,yeh76,joannopoulos97,knight03,russell03,almeida04,weeber03,yariv99,lin14prx,cohen2004,alberucci13}. Here we disclose a concept of dielectric waveguide which exploits vectorial spin-orbit interactions of light and the resulting geometric phases \cite{bliokh15,cardano15,chiao86,haldane86,berry87,bhandari97}. The approach relies on the use of anisotropic media with an optic axis that lies orthogonal to the propagation direction but is spatially modulated, so that the refractive index remains constant everywhere. A spin-controlled cumulative phase distortion is imposed on the beam, balancing diffraction for a specific polarization. Besides theoretical analysis we present an experimental demonstration of the guiding using a series of discrete geometric-phase lenses made from liquid-crystal. Our findings show that geometric phases may determine the optical guiding behaviour well beyond a Rayleigh length, paving the way to a new class of photonic devices. The concept is applicable to the whole electromagnetic spectrum.}\\[1 EM]
{\color{blue}
\href{http://dx.doi.org/10.1038/NPHOTON.2016.138}{Published in Nature Photonics, vol. 10, 571-575 (2016). DOI: 10.1038/NPHOTON.2016.138}}
\end{abstract}
\maketitle

Waveguides are central to modern photonics and optical communications. Besides the standard optical fibres --based on total internal reflection (TIR) and graded-index (GRIN) refractive potential-- and the hollow-metal-pipes for microwaves \cite{book_snyder_1983}, several more complex structures have been investigated, ranging from photonic-bandgap systems \cite{yeh76,joannopoulos97,knight03,russell03} to ``slot'' waveguides \cite{almeida04}, plasmonic waveguides \cite{weeber03}, coupled-resonators \cite{yariv99,lin14prx}, grating-mediated \cite{cohen2004} and Kapitza-effect waveguides \cite{alberucci13}.
Despite such variety, all light-guiding mechanisms investigated hitherto rely on variations, sudden or gradual, of the refractive index or -generally-  the dielectric permittivity. Even when anisotropic materials are employed to realize waveguides, as for example in liquid crystals \cite{peccianti04}, light confinement is based on the transverse modulation of the refractive index experienced by extraordinary waves through the nonuniform orientation of the optic axis with respect to the wave-vector. A fundamental question is whether the guided propagation of light can be achieved at all in structures without perturbations of the refractive index. As we shall prove, this is indeed possible provided that the transverse trapping is purely based on vectorial effects, that is, it relies on spin-orbit interactions between wave propagation and polarization states of light \cite{bliokh15,cardano15}: otherwise stated, an entirely new mechanism for light confinement.

Spin-orbit photonic interactions are strictly related to geometric Berry phases \cite{chiao86,haldane86,berry87,bhandari97}. In the context of optics, the latter are phase retardations linked exclusively to the geometry of the transformations imposed to light by the medium and independent of the optical path length \cite{bliokh15}. This concept has been already implemented in optical elements with various architectures, including patterned dielectric gratings, liquid crystals and metasurfaces \cite{bomzon01a,marrucci06a,slussarenko11,yu14,lin14science}. These devices exploit the medium anisotropy to modulate the  polarization state of light in a space-varying manner across the plane transverse to propagation. This, in turn, gives rise to a spatially inhomogeneous Pancharatnam-Berry (PB) phase \cite{pancharatnam56,berry87}, resulting in a reshaped optical wavefront. Hence, a PB optical element (PBOE) behaves as a phase mask, despite exhibiting constant ordinary and extraordinary refractive indices and a transversely-uniform optical path length, that is a flat geometry. The PB geometric phase should not be confused with the Rytov-Vladimirskii-Berry (RVB) geometric phase, or ``spin-redirection'' phase, which can also affect light propagation by inducing an additional spin-dependent spatial shift in optical media that present a transverse gradient of (isotropic) dielectric permittivity; the RVB phase is at the core of the optical Magnus effect and the spin Hall effect of light \cite{liberman92,bliokh04}.

In this work we disclose the possibility of transversely confining electromagnetic waves (and in particular light) --as in a waveguide-- by exploiting PB phases to continuously compensate diffraction in an extended continuous medium supporting beam propagation over several Rayleigh distances. As we will show, at variance with PBOEs this requires birefringent materials whose optic axis is modulated both in the transverse plane and in the longitudinal coordinate along the propagation direction, a medium structured in three-dimensions (3D).
\begin{figure}[t!]
\includegraphics[width=0.48\textwidth]{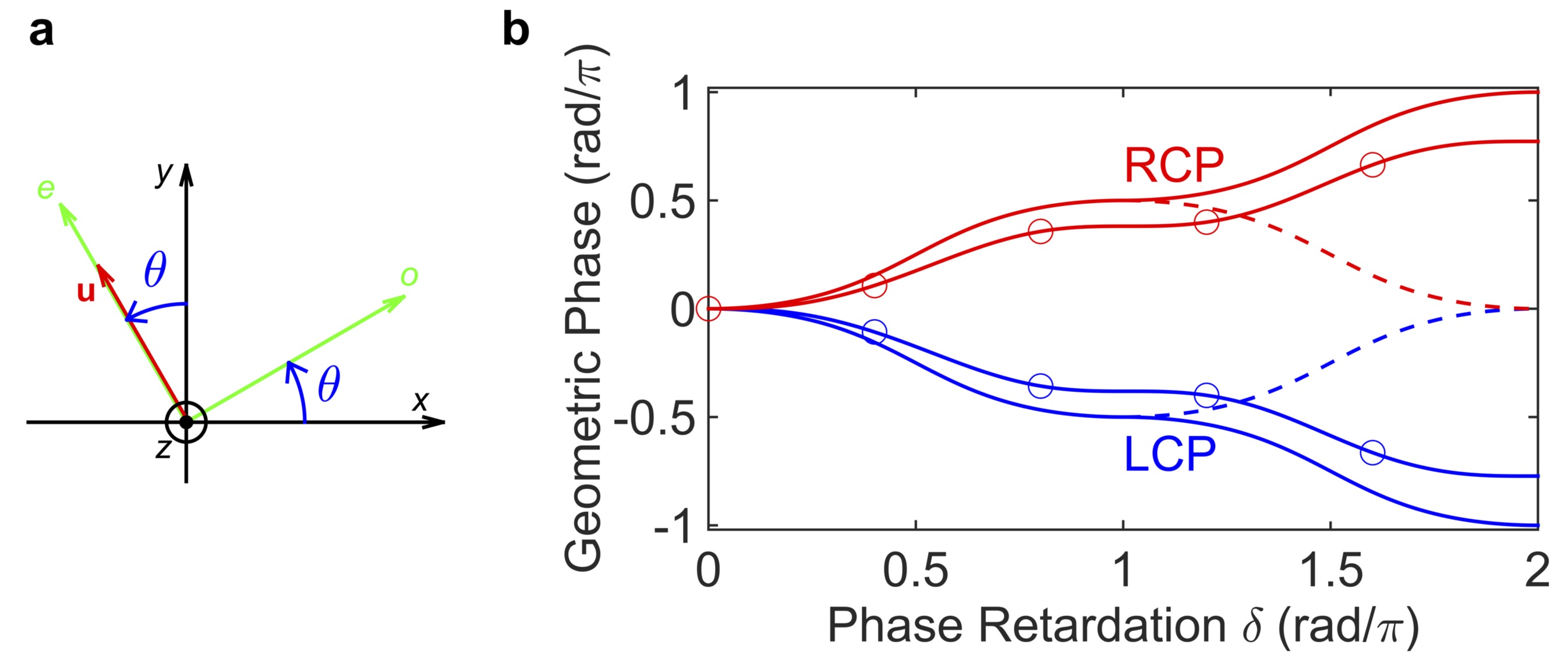}
\caption{\label{fig:geometry}{\bf Geometric-phase.} {\bf a}, Reference system $xyz$ with orientation of the optic axis $\bm{u}$ and corresponding ordinary/extraordinary ($o/e$) field directions; the angle $\theta$ between $\bm{u}$ and the axis $y$ varies from point to point. {\bf b}, Geometric phase acquired by a plane wave, CP at the input, propagating along $z$ in a transversely homogeneous medium with $\theta=\pi/4$ as a function of the birefringence retardation $\delta(z)$, relative to the case with $\theta=0$. The geometric phase sign is fixed by the CP input handedness (blue and red lines). If $\theta$ is uniform along $z$ (dashed lines), the geometric phase reaches a maximum (in the example $\pi/2$) when $\delta=\pi$ and then decreases to zero for $\delta=2\pi$. If the angle $\theta$ is suddenly inverted at $\delta=\pi$ (solid lines), the phase grows monotonically. If $\theta$ is sinusoidally modulated along $z$ (solid lines with circles), the phase increases monotonically at a slightly lower rate than in the previous case.}
\end{figure}
\begin{figure*}[t!]
\includegraphics[width=0.9\textwidth]{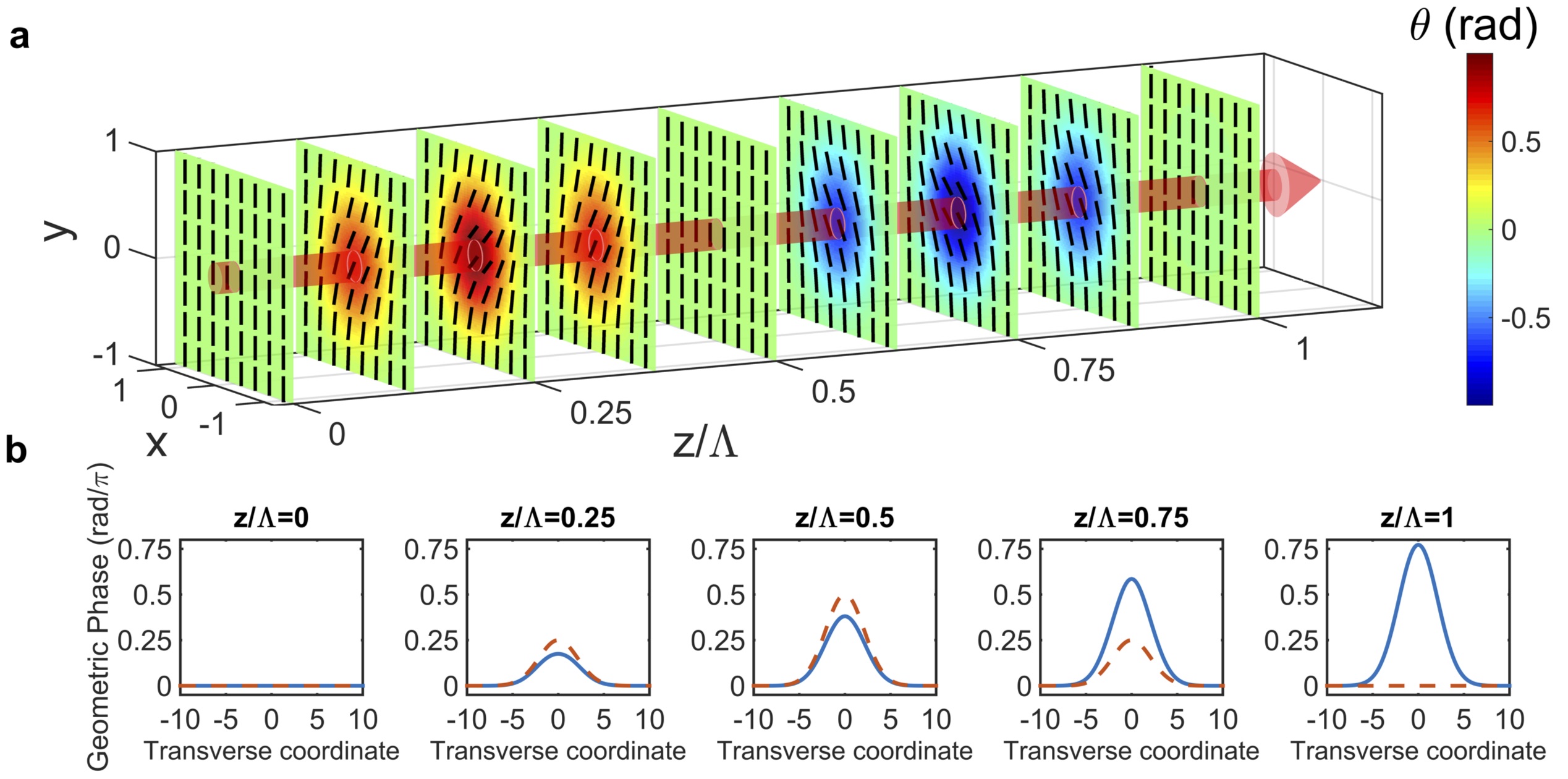}
\caption{\label{fig:waveguide}{\bf Concept of geometric-phase waveguide.} {\bf a}, 3D illustration of a continuously modulated geometric-phase waveguide: the orientation of the optic axis is longitudinally sinusoidal and transversely Gaussian. We sketch nine sections within a modulation period, with the black rods representing the optic axis and the colours corresponding to $\theta$; the guided light beam is represented as a red arrow. {\bf b}, Geometric phase accumulation across the beam profile versus propagation in the plane wave limit (that is, without diffraction), corresponding to {\bf a} (blue solid line) and in the limit of an optic axis that is unmodulated along $z$ (red dashed lines). Here the maximum $\theta$ is $\pi/4$.}
\end{figure*}
%

We investigate light propagation along the axis $z$ of an inhomogeneous uniaxial dielectric. The optic axis $\bm{u}$ is assumed to be space-varying, but lying everywhere in the $xy$ plane transverse to propagation. Its point-wise orientation is described by the angle $\theta(x,y,z)$ between $\bm{u}$ and the $y$ axis in the laboratory frame (see Fig.\ \ref{fig:geometry}a). We also assume that the principal values $\epsilon_{\parallel}$ and $\epsilon_{\perp}$ of the (relative) permittivity tensor are uniform, corresponding to constant refractive indices $n_o=\sqrt{\epsilon_{\perp}}$ and $n_e=\sqrt{\epsilon_{\parallel}}$ for ordinary and extraordinary eigenwaves, respectively. Without loss of generality, we take $n_e > n_o$.

Let us first recall that, for plane waves propagating with wave-vector along $z$ in a homogeneous uniaxial medium, the ordinary and extraordinary eigenfields have amplitudes $\psi_o(z)=e^{ik_0 n_o z}\psi_o(0)$ and $\psi_e(z)=e^{ik_0 n_e z}\psi_e(0)$, respectively, with $k_0=2\pi/\lambda$ the vacuum wavenumber and $\lambda$ the wavelength. In other words, the two waves propagate independently of each other and acquire a relative phase retardation $\delta(z)=k_0 \Delta n z$ versus propagation, where $\Delta n = n_e - n_o$ is the birefringence. If we now turn from the usual ordinary/extraordinary linear polarization basis to the left/right (L/R) circular polarization (CP) basis, the same evolution for LCP/RCP wave amplitudes is described by (see Methods for a derivation):
\begin{eqnarray}
\psi_L(z) &=& e^{i\bar{n}k_0z} \left[\cos\left(\frac{\delta}{2}\right) \psi_L(0) - i\sin\left(\frac{\delta}{2}\right)e^{i2\theta}\psi_R(0)\right] \nonumber\\
\\
\psi_R(z) &=& e^{i\bar{n}k_0z} \left[\cos\left(\frac{\delta}{2}\right) \psi_R(0) - i\sin\left(\frac{\delta}{2}\right)e^{-i2\theta}\psi_L(0)\right], \nonumber
\label{eqpboe}
\end{eqnarray}
where $\bar{n}=(n_o+n_e)/2$ is the average refractive index. Equations \eqref{eqpboe} point out that the two forward-propagating circular waves evolve with a common phase  $\bar{n}k_0z$ and, in addition, periodically exchange handedness (that is LCP becomes RCP and vice versa) acquiring an additional phase factor $\pm 2\theta$ ($+/-$ for initial RCP/LCP, respectively). This extra phase clearly has a geometric nature and is an example of PB phase \cite{pancharatnam56,berry87,bhandari97}. It should be also noted that this phase does not arise from modulations of the ordinary and extraordinary refractive indices, which are constant in the medium, nor is associated with net energy exchanges between the ordinary and extraordinary linearly-polarized wave components, which maintain their respective amplitudes all along. The two CP waves completely interchange after a propagation distance $z_\mathrm{coh}/2$ such that $\delta(z_\mathrm{coh}/2)=\pi$,  then the process reverts and the optical field retrieves its initial state in $z=z_\mathrm{coh}$ where $\delta(z_\mathrm{coh})=2\pi$ (see Fig.\ \ref{fig:geometry}b). Hence, the geometric phase in such uniform medium oscillates but does not accumulate over distance. While in PBOEs the propagation can be halted when $\delta(z)=\pi$ by properly arranging the medium length (or its birefringence), 
so as to obtain a non vanishing PB phase at the output, in a system with extended propagation length (as a waveguide) the described geometric phase appears to play no significant role.
\begin{figure*}[ht!]
\includegraphics[width=0.94\textwidth]{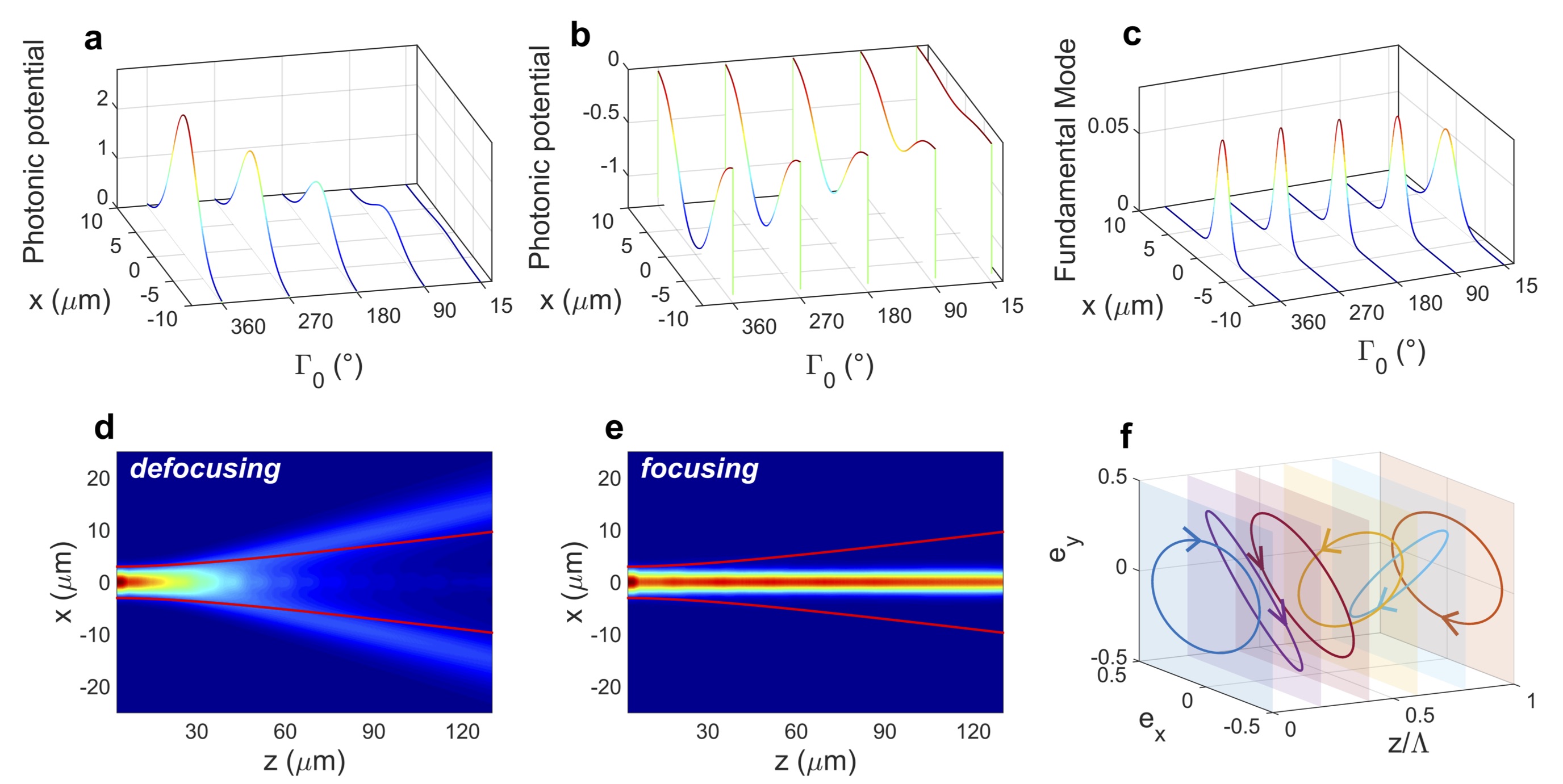}
\caption{{\label{fig:simulations} {\bf Theory and simulations.} {\bf a-b}, Effective photonic potential $V(x)2\bar{n}/k_0$ versus $x$ and maximum $\theta$ angle $\Gamma_0$ (we assumed a Gaussian distribution for the transverse orientation by setting $\Gamma=\Gamma_0 \exp{(-x^2/w_D^2)}$) perceived by the defocused ({\bf a},  LCP input) and the confined ({\bf b}, RCP input) waves, respectively. The terms proportional to $\Gamma^2$ are accounted for (see Methods). {\bf c}, Corresponding fundamental guided mode; represented is the field amplitude versus $x$ and $\Gamma_0$. {\bf d-f}, FDTD simulations for $\Gamma_0=15^\circ$ when the input beam is LCP ({\bf d}) and RCP ({\bf e}), respectively; the color scale gives the local light intensity; the red lines give the beam radius evolution for a homogeneous medium, that is for ordinary diffraction. {\bf f}, Evolution of the confined beam polarization state within a modulation period. Here $\lambda=1$ $\mu $m, $n_o=1.5$, $n_e=1.7$, $\sigma(z)$ is sinusoidal and the transverse distribution has $w_D=5$ $\mu$m.} }
\end{figure*}

A simple equalization approach to build-up a geometric phase along $z$ consists of  introducing a periodic modulation in order to counteract its recurring cancellation, analogous to dispersion-compensation in fibres and quasi-phase-matching in nonlinear optics \cite{Lin:80, ABDP}. This requires to periodically invert the sign of $\theta$ along $z$ with the same spatial period as the ``natural'' interchange described above, so that the PB phase will keep adding up  (with the same positive or negative sign depending on the input polarization) and accumulate monotonically over distance (Fig.\ \ref{fig:geometry}b). That is, we must have $\theta(z)=\theta(z+\Lambda)$ where $\Lambda=z_\mathrm{coh}=2\pi/(\Delta n k_0)=\lambda/\Delta n$ (the average value of $\theta(z)$, even if nonzero, plays no role). The resulting PB phase can then be exploited to control light over an extended propagation length. In particular, to achieve light confinement the phase retardation needs to be larger on the beam axis than in the outer regions, as in TIR or GRIN optical fibres. Such a phase modulation across the beam gives rise to a focusing effect able to counteract the natural diffraction and leading to transverse confinement and guidance. By combining longitudinal ($z$) and transverse  ($xy$ plane) modulations of $\theta$, an overall 3D structure described by $\theta(x,y,z)=\sigma(z)\Gamma(x,y)$ is obtained, with $\sigma(z)=\sigma(z+\Lambda)$ a periodic function to yield a monotonic growth of the geometric phase 
and $\Gamma(x,y)$ a transverse profile which defines the waveguide cross-section. A sample sketch of such a structure is in Fig.\ \ref{fig:waveguide}a, whereas Fig.~\ref{fig:waveguide}b shows the corresponding geometric phase accumulation in the plane-wave limit. We stress that in this inhomogeneous anisotropic medium the optic axis $\bm{u}$ remains always orthogonal to the propagation direction, so no changes to the ordinary or extraordinary refractive indices may contribute to guiding. Moreover, it will be shown that, within the validity domain of our approximations, no energy exchange between ordinary and extraordinary polarization components takes place, so that no variation of the average refractive index can contribute either.

We developed a full analytic theory of the afore described PB guiding mechanism in the frame of the slowly-varying-envelope and small-anisotropy approximations (see Methods). The main results, in a simpler geometry with one transverse coordinate $x$ for the sake of simplicity and assuming a sinusoidal $z$-modulation for $\theta$, can be summarized in a dynamical equation for the wave amplitude $A$ corresponding to a given CP input:
\begin{equation}
i\frac{\partial A}{\partial z}=-\frac{1}{2\bar{n} k_0}\frac{\partial^2 A}{\partial x^2} + V(x) A
\label{helmholtz}
\end{equation}
where $V(x)\approx \pm(\pi/\Lambda)\Gamma(x)$ with $+/-$ for input LCP/RCP, respectively. Higher-order small corrections to $V(x)$ have been omitted for simplicity (see Methods for the complete expression). Equation\ (\ref{helmholtz}) is fully equivalent to a 1D Schr\"odinger equation for a particle oscillating in a potential $V(x)$ (with $z$ playing the role of time) or to the standard (paraxial) Helmholtz equation for light propagating in a GRIN medium with refractive index $n(x)$ such that $V(x)=-k_0[n^2(x)-\bar{n}^2]/(2\bar{n})$. Depending on the sign of $\Gamma(x)$, either LCP or RCP perceive a trapping potential and get confined, while the orthogonal CP undergoes defocusing and diffracts even faster than normal, confirming once again the spin-orbit nature of this interaction. For the confined CP component the structure behaves analogously to a standard GRIN waveguide with index profile $n(x)$ (the CP state refers to the input, as the circular polarization continuously evolves between left and right during propagation, see Fig.\ \ref{fig:simulations}f). Examples of the calculated effective potential and corresponding guided modes are given in Fig.\ \ref{fig:simulations}a-c. In order to check the validity of our theory we carried out finite-difference time-domain (FDTD) numerical simulations of light propagation in the waveguide structure, solving the full Maxwell equations in the space-variant birefringent medium. Figure\ \ref{fig:simulations}d-f provides examples of the obtained results, in excellent agreement with the theoretical predictions.

\begin{figure}[t!]
\includegraphics[width=0.48\textwidth]{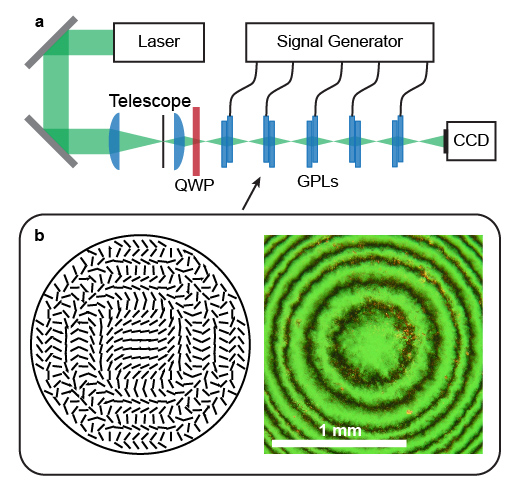}
\caption{\label{fig:apparatus} {\bf Apparatus.} {\bf a}, Experimental setup: five equally-spaced electrically-tuned GPLs form a discrete-element geometric-phase waveguide. A 532 nm continuous-wave Gaussian beam is adjusted in transverse size with a telescope for matching the fundamental mode of the waveguide. It is then circularly polarized with a quarter-wave plate (QWP) and launched into the waveguide. Beam profiles at various intermediate positions $z$ along the propagation and at the output were acquired by a movable CCD camera and used to reconstruct the mode parameters of the beam after each GPL. {\bf b}, Distribution of the optic axis and corresponding photograph of a GPL imaged between crossed polarizers; dark fringes correspond to regions where the optic axis is aligned parallel to one of the polarizers.}
\end{figure}

We demonstrated this novel approach to guiding light with a proof-of-principle experiment. Rather than using the continuous structure described above, we realized a simpler ``discrete-element'' PB-waveguide consisting of equally spaced PBOEs alternating with a homogeneous isotropic dielectric (air); each PBOE is essentially equivalent to a thin slice of the PB-waveguide and acts as a focusing element, that is, a geometric-phase lens (GPL) \cite{bhandari97,hasman03,roux06,marrucci06a}. In other words, we mimicked the operation of the PB-waveguide with a sequence of equally spaced converging lenses, exclusively using PB phases (Fig.\ \ref{fig:apparatus}a).
\begin{figure*}[t!]
\includegraphics[width=0.95\textwidth]{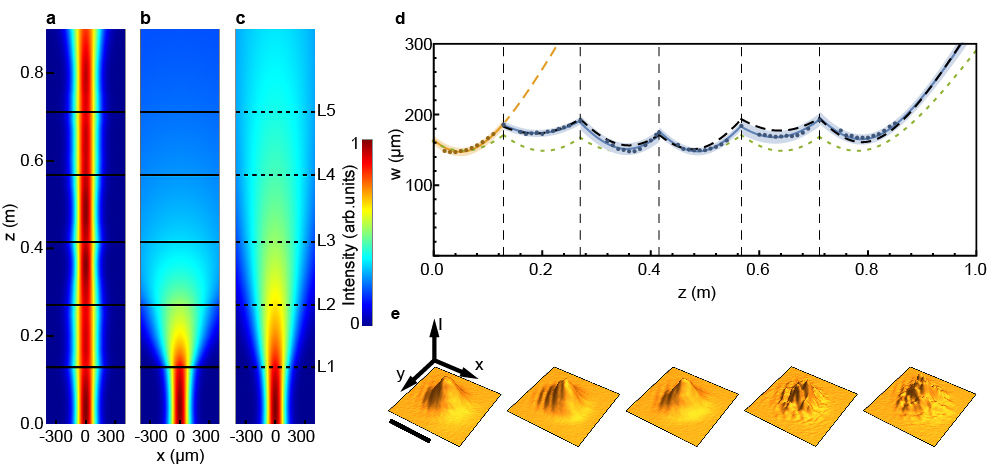}
\caption{\label{fig:experiment} {\bf Experiment.} {\bf a}-{\bf c}, Data-reconstructed beam evolution for the following cases: {\bf a}, guided mode, obtained for RCP input; {\bf b}, divergent beam, obtained for LCP input; {\bf c}, free-space diffracting beam for the same input parameters. The color scale gives the local light intensity. Horizontal lines and L1-L5 labels indicate the GPL positions within the discrete sequence (dashed lines mark removed GPLs). {\bf d}, Beam radius versus $z$ in the guided case. Dots are the measurement data, blue solid lines are the corresponding Gaussian-beam fits between subsequent GPLs, with blue shaded areas indicating confidence regions at one standard deviation; black dashed line is the theoretical prediction from the ABCD method, accounting for the Gaussian-beam imperfections (quantified by the $M^2$ parameter); dashed green line corresponds to the ideal Gaussian beam case, with $M^2=1$. Vertical dashed lines mark the GPL positions. {\bf e}, Measured light intensity profiles ($I$ versus $x,y$) at the input plane of each GPL for the guided case; scale-bar corresponds to 400 $\mu$m.}
\end{figure*}

Our GPLs were thin films of a birefringent uniaxial, nematic liquid crystals, having a transverse distribution of the optic axis given by $\theta=\alpha(x^2+y^2)$, with $\alpha$ a constant, see Fig.\ \ref{fig:apparatus}b. Neglecting diffraction within the finite GPL thickness, the action of each lens on a CP input beam is described by Eqs.\ (\ref{eqpboe}). For $\delta=\pi$, the polarization handedness is inverted at the output and the outgoing wave acquires a geometric phase $\pm2\theta(x,y)$, which is equivalent to the phase of a thin lens having focal length $f=\pm\frac{\pi}{2\alpha\lambda}$. Hence, the GPL acts as a focusing element for one CP handedness but defocusing for the opposite one. Since the circular polarization handedness is inverted at each GPL, in order to balance out diffraction throughout the structure, we flipped the sign of $\alpha$ at each step, resulting in a longitudinal $\theta$ oscillation as for the continuous PB waveguide case. The fundamental mode of our discrete waveguide, the shape-preserving Gaussian beam that propagates with period equal to the distance $d$ between opposite lenses, has a beam waist $w_0=\sqrt{\lambda[(4f-d)d]^{1/2}/(2n_i\pi)}$ centred between subsequent GPLs, where $n_i$ is the refractive index of the isotropic medium between elements.

For our experiments, we set up a sequence of five GPLs, rotating every other one by $\pi$ around the $y$ axis so as to produce alternating signs of $\alpha$; moreover, exploiting the electro-optic response of nematic liquid crystals, the lenses were electrically tuned to $\delta=\pi$ (see Methods for details). We characterized light propagation in the structure for both LCP and RCP inputs and compared it with free-space propagation. As can be seen in Figs.~\ref{fig:experiment}a-c, the experimental results are in very good agreement with the polarization-dependent waveguiding predicted for the continuous case: only one input CP handedness was confined in the PB-waveguide with a shape-preserving mode, whereas the opposite polarization was radiated off almost immediately. The acquired data matched well the theory, as shown in Fig.\ \ref{fig:experiment}d.

In conclusion, we have shown, both theoretically and experimentally, that geometric Pancharatnam-Berry phases can be used for transverse confinement of electromagnetic waves, thus introducing an entirely new light-guiding principle that exploits spin-orbit optical interactions and the vectorial nature of electromagnetic radiation. Besides its fundamental interest, the proposed approach is technologically relevant for future integrated optics systems, including those involving metasurfaces \cite{yu14,lin14science}. The development of novel generations of PB guided-wave photonics and manipulation of light is envisioned in dielectrics and metamaterials for the whole spectrum, from terahertz to ultraviolet. 

\bibliographystyle{naturemag_noURLmod}
\bibliography{geolightguide}

\begin{thebibliography}{10}
\expandafter\ifx\csname url\endcsname\relax
  \def\url#1{\texttt{#1}}\fi
\expandafter\ifx\csname urlprefix\endcsname\relax\def\urlprefix{URL }\fi
\providecommand{\bibinfo}[2]{#2}
\providecommand{\eprint}[2][]{\url{#2}}

\bibitem{book_snyder_1983}
\bibinfo{author}{Snyder, A.W.} \& \bibinfo{author}{Love, J.D.}
\newblock \emph{\bibinfo{title}{Optical Waveguide Theory}}
  (\bibinfo{publisher}{Chapman and Hall}, \bibinfo{address}{New York},
  \bibinfo{year}{1983}).

\bibitem{yeh76}
\bibinfo{author}{Yeh, P.} \& \bibinfo{author}{Yariv, A.}
\newblock \bibinfo{title}{Bragg reflection waveguides}.
\newblock \emph{\bibinfo{journal}{Opt. Commun.}} \textbf{\bibinfo{volume}{19}},
  \bibinfo{pages}{427--430} (\bibinfo{year}{1976}).

\bibitem{joannopoulos97}
\bibinfo{author}{Joannopoulos, J.D.}, \bibinfo{author}{Villeneuve, P.R.} \&
  \bibinfo{author}{Fan, S.H.}
\newblock \bibinfo{title}{Photonic crystals: Putting a new twist on light}.
\newblock \emph{\bibinfo{journal}{Nature}} \textbf{\bibinfo{volume}{386}},
  \bibinfo{pages}{143--149} (\bibinfo{year}{1997}).

\bibitem{knight03}
\bibinfo{author}{Knight, J.C.}
\newblock \bibinfo{title}{Photonic crystal fibres}.
\newblock \emph{\bibinfo{journal}{Nature}} \textbf{\bibinfo{volume}{424}},
  \bibinfo{pages}{847--851} (\bibinfo{year}{2003}).

\bibitem{russell03}
\bibinfo{author}{Russell, P.}
\newblock \bibinfo{title}{Photonic crystal fibers}.
\newblock \emph{\bibinfo{journal}{Science}} \textbf{\bibinfo{volume}{299}},
  \bibinfo{pages}{358--362} (\bibinfo{year}{2003}).

\bibitem{almeida04}
\bibinfo{author}{Almeida, V.R.}, \bibinfo{author}{Xu, Q.},
  \bibinfo{author}{Barrios, C.A.} \& \bibinfo{author}{Lipson, M.}
\newblock \bibinfo{title}{Guiding and confining light in void nanostructure}.
\newblock \emph{\bibinfo{journal}{Opt. Lett.}} \textbf{\bibinfo{volume}{29}},
  \bibinfo{pages}{1209--1211} (\bibinfo{year}{2004}).

\bibitem{weeber03}
\bibinfo{author}{Weeber, J.C.}, \bibinfo{author}{Lacroute, Y.} \&
  \bibinfo{author}{Dereux, A.}
\newblock \bibinfo{title}{Optical near-field distributions of surface plasmon
  waveguide modes}.
\newblock \emph{\bibinfo{journal}{Phys. Rev. B}} \textbf{\bibinfo{volume}{68}},
  \bibinfo{pages}{115401} (\bibinfo{year}{2003}).

\bibitem{yariv99}
\bibinfo{author}{Yariv, A.}, \bibinfo{author}{Xu, Y.}, \bibinfo{author}{Lee,
  R.K.} \& \bibinfo{author}{Scherer, A.}
\newblock \bibinfo{title}{Coupled-resonator optical waveguide: a proposal and
  analysis}.
\newblock \emph{\bibinfo{journal}{Opt. Lett.}} \textbf{\bibinfo{volume}{24}},
  \bibinfo{pages}{711--713} (\bibinfo{year}{1999}).

\bibitem{lin14prx}
\bibinfo{author}{Lin, Q.} \& \bibinfo{author}{Fan, S.}
\newblock \bibinfo{title}{Light guiding by effective gauge field for photons}.
\newblock \emph{\bibinfo{journal}{Phys. Rev. X}} \textbf{\bibinfo{volume}{4}},
  \bibinfo{pages}{031031} (\bibinfo{year}{2014}).

\bibitem{cohen2004}
\bibinfo{author}{Cohen, O.}, \bibinfo{author}{Freedman, B.},
  \bibinfo{author}{Fleischer, J.W.}, \bibinfo{author}{Segev, M.} \&
  \bibinfo{author}{Christodoulides, D.N.}
\newblock \bibinfo{title}{Grating-mediated waveguiding}.
\newblock \emph{\bibinfo{journal}{Phys. Rev. Lett.}}
  \textbf{\bibinfo{volume}{93}}, \bibinfo{pages}{103902}
  (\bibinfo{year}{2004}).

\bibitem{alberucci13}
\bibinfo{author}{Alberucci, A.}, \bibinfo{author}{Marrucci, L.} \&
  \bibinfo{author}{Assanto, G.}
\newblock \bibinfo{title}{Light confinement via periodic modulation of the
  refractive index}.
\newblock \emph{\bibinfo{journal}{New J. Phys.}} \textbf{\bibinfo{volume}{15}},
  \bibinfo{pages}{083013} (\bibinfo{year}{2013}).

\bibitem{bliokh15}
\bibinfo{author}{Bliokh, K.Y.}, \bibinfo{author}{Rodriguez-Fortuno, F.J.},
  \bibinfo{author}{Nori, F.} \& \bibinfo{author}{Zayats, A.V.}
\newblock \bibinfo{title}{Spin-orbit interactions of light}.
\newblock \emph{\bibinfo{journal}{Nature Photon.}}
  \textbf{\bibinfo{volume}{9}}, \bibinfo{pages}{796--808}
  (\bibinfo{year}{2015}).

\bibitem{cardano15}
\bibinfo{author}{Cardano, F.} \& \bibinfo{author}{Marrucci, L.}
\newblock \bibinfo{title}{Spin-orbit photonics}.
\newblock \emph{\bibinfo{journal}{Nature Photon.}}
  \textbf{\bibinfo{volume}{9}}, \bibinfo{pages}{776--778}
  (\bibinfo{year}{2015}).

\bibitem{chiao86}
\bibinfo{author}{Chiao, R.Y.} \& \bibinfo{author}{Wu, Y.S.}
\newblock \bibinfo{title}{Manifestations of {Berry}'s topological phase for the
  photon}.
\newblock \emph{\bibinfo{journal}{Phys. Rev. Lett.}}
  \textbf{\bibinfo{volume}{57}}, \bibinfo{pages}{933} (\bibinfo{year}{1986}).

\bibitem{haldane86}
\bibinfo{author}{Haldane, F.D.M.}
\newblock \bibinfo{title}{Path dependence of the geometric rotation of
  polarization in optical fibers}.
\newblock \emph{\bibinfo{journal}{Opt. Lett.}} \textbf{\bibinfo{volume}{11}},
  \bibinfo{pages}{730--732} (\bibinfo{year}{1986}).

\bibitem{berry87}
\bibinfo{author}{Berry, M.V.}
\newblock \bibinfo{title}{The adiabatic phase and {Pancharatnam}'s phase for
  polarized light}.
\newblock \emph{\bibinfo{journal}{J. Mod. Opt.}} \textbf{\bibinfo{volume}{34}},
  \bibinfo{pages}{1401--1407} (\bibinfo{year}{1987}).

\bibitem{bhandari97}
\bibinfo{author}{Bhandari, R.}
\newblock \bibinfo{title}{Polarization of light and topological phases}.
\newblock \emph{\bibinfo{journal}{Phys. Rep.}} \textbf{\bibinfo{volume}{281}},
  \bibinfo{pages}{1--64} (\bibinfo{year}{1997}).

\bibitem{peccianti04}
\bibinfo{author}{Peccianti, M.}, \bibinfo{author}{Conti, C.},
  \bibinfo{author}{Assanto, G.}, \bibinfo{author}{Luca, A.D.} \&
  \bibinfo{author}{Umeton, C.}
\newblock \bibinfo{title}{Routing of anisotropic spatial solitons and
  modulational instability in liquid crystals}.
\newblock \emph{\bibinfo{journal}{Nature}} \textbf{\bibinfo{volume}{432}},
  \bibinfo{pages}{733--737} (\bibinfo{year}{2004}).

\bibitem{bomzon01a}
\bibinfo{author}{Bomzon, Z.}, \bibinfo{author}{Kleiner, V.} \&
  \bibinfo{author}{Hasman, E.}
\newblock \bibinfo{title}{{Pancharatnam}-{Berry} phase in space-variant
  polarization-state manipulations with subwavelength gratings}.
\newblock \emph{\bibinfo{journal}{Opt. Lett.}} \textbf{\bibinfo{volume}{26}},
  \bibinfo{pages}{1424--1426} (\bibinfo{year}{2001}).

\bibitem{marrucci06a}
\bibinfo{author}{Marrucci, L.}, \bibinfo{author}{Manzo, C.} \&
  \bibinfo{author}{Paparo, D.}
\newblock \bibinfo{title}{{Pancharatnam}-{Berry} phase optical elements for
  wavefront shaping in the visible domain: switchable helical modes
  generation}.
\newblock \emph{\bibinfo{journal}{Appl. Phys. Lett.}}
  \textbf{\bibinfo{volume}{88}}, \bibinfo{pages}{221102}
  (\bibinfo{year}{2006}).

\bibitem{slussarenko11}
\bibinfo{author}{Slussarenko, S.} \emph{et~al.}
\newblock \bibinfo{title}{Tunable liquid crystal q-plates with arbitrary
  topological charge}.
\newblock \emph{\bibinfo{journal}{Opt. Express}} \textbf{\bibinfo{volume}{19}},
  \bibinfo{pages}{4085--4090} (\bibinfo{year}{2011}).

\bibitem{yu14}
\bibinfo{author}{Yu, N.} \& \bibinfo{author}{Capasso, F.}
\newblock \bibinfo{title}{Flat optics with designer metasurfaces}.
\newblock \emph{\bibinfo{journal}{Nature Mater.}}
  \textbf{\bibinfo{volume}{13}}, \bibinfo{pages}{139--150}
  (\bibinfo{year}{2014}).

\bibitem{lin14science}
\bibinfo{author}{Lin, D.}, \bibinfo{author}{Fan, P.}, \bibinfo{author}{Hasman,
  E.} \& \bibinfo{author}{Brongersma, M.L.}
\newblock \bibinfo{title}{Dielectric gradient metasurface optical elements}.
\newblock \emph{\bibinfo{journal}{Science}} \textbf{\bibinfo{volume}{345}},
  \bibinfo{pages}{298--302} (\bibinfo{year}{2014}).

\bibitem{pancharatnam56}
\bibinfo{author}{Pancharatnam, S.}
\newblock \bibinfo{title}{Generalized theory of interference, and its
  applications}.
\newblock \emph{\bibinfo{journal}{Proc. Indian Acad. Sci. A}}
  \textbf{\bibinfo{volume}{44}}, \bibinfo{pages}{0370--0089}
  (\bibinfo{year}{1956}).

\bibitem{liberman92}
\bibinfo{author}{Liberman, V.S.} \& \bibinfo{author}{Zel'dovich, B.Y.}
\newblock \bibinfo{title}{Spin-orbit interaction of a photon in an
  inhomogeneous medium}.
\newblock \emph{\bibinfo{journal}{Phys. Rev. A}} \textbf{\bibinfo{volume}{46}},
  \bibinfo{pages}{5199--5207} (\bibinfo{year}{1992}).

\bibitem{bliokh04}
\bibinfo{author}{Bliokh, K.Y.} \& \bibinfo{author}{Bliokh, Y.P.}
\newblock \bibinfo{title}{Modified geometrical optics of a smoothly
  inhomogeneous isotropic medium: The anisotropy, {B}erry phase, and the
  optical {M}agnus effect}.
\newblock \emph{\bibinfo{journal}{Phys. Rev. E}} \textbf{\bibinfo{volume}{70}},
  \bibinfo{pages}{026605} (\bibinfo{year}{2004}).

\bibitem{Lin:80}
\bibinfo{author}{Lin, C.}, \bibinfo{author}{Cohen, L.G.} \&
  \bibinfo{author}{Kogelnik, H.}
\newblock \bibinfo{title}{Optical-pulse equalization of low-dispersion
  transmission in single-mode fibers in the 1.3--1.7-$\mu$m spectral region}.
\newblock \emph{\bibinfo{journal}{Opt. Lett.}} \textbf{\bibinfo{volume}{5}},
  \bibinfo{pages}{476--478} (\bibinfo{year}{1980}).

\bibitem{ABDP}
\bibinfo{author}{Armstrong, J.A.}, \bibinfo{author}{Bloembergen, N.},
  \bibinfo{author}{Ducuing, J.} \& \bibinfo{author}{Pershan, P.S.}
\newblock \bibinfo{title}{Interactions between light waves in a nonlinear
  dielectric}.
\newblock \emph{\bibinfo{journal}{Phys. Rev.}} \textbf{\bibinfo{volume}{127}},
  \bibinfo{pages}{1918--1939} (\bibinfo{year}{1962}).

\bibitem{hasman03}
\bibinfo{author}{Hasman, E.}, \bibinfo{author}{Kleiner, V.},
  \bibinfo{author}{Biener, G.} \& \bibinfo{author}{Niv, A.}
\newblock \bibinfo{title}{Polarization dependent focusing lens by use of
  quantized {Pancharatnam}-{Berry} phase diffractive optics}.
\newblock \emph{\bibinfo{journal}{Appl. Phys. Lett.}}
  \textbf{\bibinfo{volume}{82}}, \bibinfo{pages}{328--330}
  (\bibinfo{year}{2003}).

\bibitem{roux06}
\bibinfo{author}{Roux, F.S.}
\newblock \bibinfo{title}{Geometric phase lens}.
\newblock \emph{\bibinfo{journal}{J. Opt. Soc. Am. A}}
  \textbf{\bibinfo{volume}{23}}, \bibinfo{pages}{476--482}
  (\bibinfo{year}{2006}).

\setcounter{firstbib}{\value{NAT@ctr}}
\end{thebibliography}

\subsection*{References}
\begin{thebibliography}{10}
\setcounter{NAT@ctr}{\value{firstbib}}

\bibitem{berry94}
\bibinfo{author}{Berry, M.V.}
\newblock \bibinfo{title}{{Pancharatnam}, virtuoso of the {P}oincar\'e sphere:
  an appreciation}.
\newblock \emph{\bibinfo{journal}{Current Science}}
  \textbf{\bibinfo{volume}{67}}, \bibinfo{pages}{220--223}
  (\bibinfo{year}{1994}).

\bibitem{OskooiRo10}
\bibinfo{author}{Oskooi, A.F.} \emph{et~al.}
\newblock \bibinfo{title}{{MEEP}: A flexible free-software package for
  electromagnetic simulations by the {FDTD} method}.
\newblock \emph{\bibinfo{journal}{Computer Phys. Commun.}}
  \textbf{\bibinfo{volume}{181}}, \bibinfo{pages}{687--702}
  (\bibinfo{year}{2010}).

\bibitem{marrucci06}
\bibinfo{author}{Marrucci, L.}, \bibinfo{author}{Manzo, C.} \&
  \bibinfo{author}{Paparo, D.}
\newblock \bibinfo{title}{Optical spin-to-orbital angular momentum conversion
  in inhomogeneous anisotropic media}.
\newblock \emph{\bibinfo{journal}{Phys. Rev. Lett.}}
  \textbf{\bibinfo{volume}{96}}, \bibinfo{pages}{163905}
  (\bibinfo{year}{2006}).

\bibitem{tabiryan09}
\bibinfo{author}{Nersisyan, S.}, \bibinfo{author}{Tabiryan, N.},
  \bibinfo{author}{Steeves, D.M.} \& \bibinfo{author}{Kimball, B.R.}
\newblock \bibinfo{title}{Fabrication of liquid crystal polymer axial
  waveplates for {UV}-{IR} wavelengths}.
\newblock \emph{\bibinfo{journal}{Opt. Express}} \textbf{\bibinfo{volume}{17}},
  \bibinfo{pages}{11926--11934} (\bibinfo{year}{2009}).

\bibitem{alexeyev12}
\bibinfo{author}{Alexeyev, C.N.}
\newblock \bibinfo{title}{Circular array of anisotropic fibers: A discrete
  analog of a $q$ plate}.
\newblock \emph{\bibinfo{journal}{Phys. Rev. A}} \textbf{\bibinfo{volume}{86}},
  \bibinfo{pages}{063830} (\bibinfo{year}{2012}).

\bibitem{book_chigrinov_PA}
\bibinfo{author}{Chigrinov, V.G.}, \bibinfo{author}{Kozenkov, V.M.} \&
  \bibinfo{author}{Kwok, H.S.}
\newblock \emph{\bibinfo{title}{Photoalignment of Liquid Crystalline Materials:
  Physics and Applications}} (\bibinfo{publisher}{Wiley Publishing},
  \bibinfo{year}{2008}).

\bibitem{piccirillo10}
\bibinfo{author}{Piccirillo, B.}, \bibinfo{author}{D'Ambrosio, V.},
  \bibinfo{author}{Slussarenko, S.}, \bibinfo{author}{Marrucci, L.} \&
  \bibinfo{author}{Santamato, E.}
\newblock \bibinfo{title}{Photon spin-to-orbital angular momentum conversion
  via an electrically tunable q-plate}.
\newblock \emph{\bibinfo{journal}{Appl. Phys. Lett.}}
  \textbf{\bibinfo{volume}{97}}, \bibinfo{pages}{241104}
  (\bibinfo{year}{2010}).

\bibitem{sun98}
\bibinfo{author}{Sun, H.}
\newblock \bibinfo{title}{Thin lens equation for a real laser beam with weak
  lens aperture truncation}.
\newblock \emph{\bibinfo{journal}{Opt. Eng.}} \textbf{\bibinfo{volume}{37}},
  \bibinfo{pages}{2906--2913} (\bibinfo{year}{1998}).

\end{thebibliography}

\onecolumngrid

\vspace{1 EM}
\section*{Methods}
\subsection*{Dynamics of circularly polarized waves through a uniform uniaxial medium}
Plane-wave light propagation along $z$ in a uniform uniaxial medium can be described by a two-component Jones vector $\bm{\psi}(z)$ (in the bra-ket notation $\left|\psi \right\rangle$), representing the complex amplitudes of two orthogonal polarizations which define the chosen basis of the representation. A $2\times 2$ evolution Jones matrix $\bm{U}(z)$ then links the vector $\bm{\psi}(0)$ at the input plane $z=0$ with the vector $\bm{\psi}(z)$ at any given distance $z$. In the $xy$ laboratory basis, the Jones vector is the same as the (complex) electric field vector $\bm{\psi}_{xy}=(E_x; E_y)$. A more convenient basis is that of the ordinary/extraordinary linear polarizations, that is $\bm{\psi}_{oe}=(E_o; E_e)$. These two bases are connected by the rotation matrix $\bm{R}_{xy}(\theta)=(\cos\theta, \sin\theta; -\sin\theta, \cos\theta)$, where $\theta$ is the angle formed by the optic axis $\bm{u}$ with the $y$ axis, that is $\bm{\psi}_{oe}=\bm{R}_{xy}\cdot\bm{\psi}_{xy}$. In this $oe$ basis the evolution matrix is diagonal, with the expression
\begin{equation}\label{UOE}
\bm{U}_{oe}(z)=
	\left( \begin{array} {cc} 
  	e^{i k_0 n_o z}  & 0 \\
  	0 & e^{i k_0 n_e z} \\ 
 	\end{array} \right).
\end{equation}
Let us now introduce a third convenient basis to represent propagation, that is the LCP/RCP circular polarization basis, hereafter denoted as $LR$. In this Letter, we adopt the notation $\bm{\psi}_L=(1; -i)/\sqrt{2}$ (LCP) and $\bm{\psi}_R=(1; +i)/\sqrt{2}$ (RCP) for the basis unit vectors, corresponding to the source-point-of-view naming convention on CP states. The matrix $\bm{P}=(1,1;-i,i)/\sqrt{2}$ can then be used to switch the Jones vector from the $LR$ basis to the $xy$ one, while the rotation operator in the $LR$ basis is diagonal and takes the form $\bm{R}_{LR}(\theta)=(e^{-i\theta}, 0; 0, e^{i\theta})$. The evolution matrix in the $LR$ basis is then given by
\begin{equation}  \label{ULR}
   \bm{U}_{LR}(z;\theta)= \bm{R}^{-1}_{LR}(\theta)\cdot \bm{P}^{-1} \cdot \bm{U}_{oe}(z) \cdot \bm{P} \cdot \bm{R}_{LR}(\theta).
\end{equation}   
A straightforward calculation leads from Eq.\ \eqref{ULR} to Eq.\ \eqref{eqpboe}. As mentioned in the main text, the resulting polarization dynamics is an oscillation with period $z_{\text{coh}}=\lambda/\Delta n$.

Let us consider the solution for a purely circular polarization at the input, e.g.\ for $\psi_R(0)=1$ and $\psi_L(0)=0$. From Eq.\ \eqref{eqpboe} we get $\psi_R(z)=e^{i\bar{n}k_0 z} \cos(\frac{\delta}{2})$ and $\psi_L(z)=-i e^{i\bar{n}k_0 z} \sin(\frac{\delta}{2}) e^{2i\theta}$. From the latter expression we can calculate the phase difference between two states corresponding to two distinct orientations of the optic axis, say $\theta_1$ and $\theta_2$, respectively. Following Pancharatnam's original concept (see Refs.\ \onlinecite{pancharatnam56,berry94}), the phase delay $\Delta \phi(\theta_1,\theta_2)$ is
\begin{equation}
  \Delta \phi(\theta_1,\theta_2)= \text{arg}\left[\left\langle \psi(\theta_1) | \psi(\theta_2) \right\rangle \right] = \text{arg}\left[ \cos^2\left(\frac{\delta}{2} \right) + \sin^2\left(\frac{\delta}{2} \right) e^{2i\left(\theta_2 -\theta_1 \right)} \right]. \label{pancharatnam_phase}
\end{equation}
Expression \eqref{pancharatnam_phase} results in the transverse phase delay plotted in Fig.\ \ref{fig:geometry}b (dashed lines) in the homogeneous limit. When the optic axis distribution is flipped at $\Lambda/2$, the phase delay can be easily obtained from Eq.\ \eqref{ULR} (solid lines in Fig.\ \ref{fig:geometry}b). When the optic axis distribution is sinusoidally modulated along $z$,  the accumulated $\Delta \phi(\theta_1,\theta_2)$ can be numerically calculated by partitioning the medium in several layers, each of them short enough to make the variations of $\theta$ negligible within each layer, as shown in Fig.\ \ref{fig:geometry}b (solid lines with circles). 

\subsection*{Spin-dependent photonic potential}
In the paraxial limit (i.e., neglecting the longitudinal field components) and for small birefringence ($\Delta n\ll1$), the Maxwell equations for the electric field $\bm{\psi}_{xy}=(E_x; E_y)$ in two dimensions (i.e., with no field evolution across $y$) can be cast as
\begin{equation} \label{eq:maxwell_equation}
 \frac{\partial^2}{\partial z^2} \left ( \begin{array} {c}
  E_x \\
  E_y \\
 \end{array} \right)  = - \frac{\partial^2}{\partial x^2} \left ( \begin{array} {c}
  E_x \\
  E_y \\
 \end{array} \right) - k_0^2 \left ( \begin{array} {cc}
  \epsilon_{xx}(x,z) & \epsilon_{xy}(x,z) \\
  \epsilon_{yx}(x,z) & \epsilon_{yy}(x,z) \\
 \end{array} \right) \left ( \begin{array} {c}
  E_x \\
  E_y \\
 \end{array} \right). 
\end{equation}
We wish to adopt now the slowly-varying-envelope approximation (SVEA) (also corresponding to the paraxial wave approximation) so as to obtain a simpler first-order partial differential equation in the evolution coordinate $z$. SVEA is usually based on the presence of two very different spatial scales for the evolution along $z$, that is, a short scale of order $\lambda$ and a long scale given by the Rayleigh length $z_R = \pi \bar{n}w^2/\lambda$, where $w$ is the smallest transverse spatial scale of the problem (the beam radius and/or the transverse spatial modulations of the medium). In our case, however, we also have an intermediate scale $\Lambda=z_{\text{coh}}=\lambda/\Delta n$, with $\lambda\ll\Lambda\ll z_R$ for typical values of $\Delta n$ and $w$. It is therefore not appropriate to apply the SVEA directly to Eq.\ \eqref{eq:maxwell_equation}, as the field components $E_x$ and $E_y$ undergo a relatively rapid evolution on scale $\Lambda$ because of birefringence. It is more convenient to switch first to the space-varying $oe$ wave basis introduced in Section A, with a  rotation by the angle $\theta(x,z)$ around the $z$ axis. This change of basis diagonalizes the effect of birefringence and hence allows adopting the SVEA in an optimal way, but at the same time it brings about the space-varying geometric phases which will affect the resulting slow-envelope dynamics. In the rotated basis, Eq.\ (\ref{eq:maxwell_equation}) becomes
\begin{eqnarray}
 &&  \frac{\partial^2 \bm{\psi}_{oe}}{\partial z^2} - 2i \frac{\partial\theta}{\partial z} \bm{S}_2 \cdot \frac{\partial \bm{\psi}_{oe}}{\partial z}   =
 -  \frac{\partial^2 \bm{\psi}_{oe}}{\partial x^2} + 2i  \frac{\partial\theta}{\partial x} \bm{S}_2 \cdot \frac{\partial \bm{\psi}_{oe}}{\partial  x} +i \left( \frac{\partial^2 \theta}{\partial x^2} + \frac{\partial^2\theta}{\partial z^2} \right) \bm{S}_2\cdot \bm{\psi}_{oe}   \nonumber \\
&& + \left[ \left(\frac{\partial\theta}{\partial x} \right)^2 + \left(\frac{\partial\theta}{\partial z}\right)^2  \right] \bm{\psi}_{oe} - k_0^2 \bm{\epsilon}_D \cdot \bm{\psi}_{oe}, 
\label{eq:maxwell_rotated_inho}
\end{eqnarray}
where we introduced the Pauli matrix $\bm{S}_2=\left(0,-i;i,0 \right)$ and the diagonalized permittivity tensor $\bm{\epsilon}_D=(\epsilon_\bot, 0 ; 0, \epsilon_\|)$. We now set $\bm{\psi}_{oe}(x,z)=\bm{U}_{oe}(z)\cdot \bm{\psi}(x,z)$, where $\bm{U}_{oe}(z)$ is the evolution matrix in the $oe$ basis for plane waves in a uniform medium, as given in Eq.\ (\ref{UOE}), and $\bm{\psi}(z)$ are slowly-varying amplitudes. Moreover, let us assume sinusoidal $z$-modulation of the optic axis, as given by $\theta(x,z)=\Gamma(x) \sin(2\pi z/\Lambda)$. We then multiply both sides of Eq.\ (\ref{eq:maxwell_rotated_inho}) by $\bm{U}_{oe}^{-1}$ from the left and take a $z$-average over a period $\Lambda$ of all terms, assuming that the amplitudes $\bm{\psi}(z)$ vary slowly enough that they can be taken out of the averaging operation (this obviously requires $\Lambda \ll z_R$). The quadratic terms in $\theta(x,z)$ will then contribute with a non-vanishing average, independent of the polarization evolution. In addition, the sinusoidal terms linear in $\theta(x,z)$ combine with the oscillations in polarization described by the $z$-evolved Pauli matrix $\tilde{\bm{S}}_2(z)=\bm{U}_{oe}^{-1}(z)\cdot\bm{S}_2\cdot\bm{U}_{oe}(z)=\bm{S}_2 \cos\left(k_0\Delta n z \right) + \bm{S}_1 \sin\left( k_0\Delta n z \right)$, where $\bm{S}_1 = (0, 1; 1, 0)$ is the first Pauli matrix, leading to other phase-matched constant terms. We thus obtain the following dynamical equation for the slow amplitudes $\bm{\psi}(z)$:
\begin{eqnarray}
&& \frac{\partial^2\bm{\psi}}{\partial z^2}+\left( 2ik_0\bm{N}-i\frac{2\pi}{\Lambda}\Gamma(x)\bm{S}_2\right)\cdot\frac{\partial\bm{\psi}}{\partial z} =
-\frac{\partial^2\bm{\psi}}{\partial x^2}-2k_0\frac{\pi}{\Lambda}\Gamma(x)\bm{S}_2\cdot\bm{N}\cdot\bm{\psi} +i \frac{\partial\Gamma}{\partial x}\bm{S}_1\cdot\frac{\partial\bm{\psi}}{\partial x}
\nonumber\\
&& +\frac{i}{2}\left[\frac{\partial^2\Gamma}{\partial x^2}-\left(\frac{2\pi}{\Lambda}\right)^2\Gamma(x)\right]\bm{S}_1\cdot\bm{\psi}
+\frac{1}{2}\left[\left(\frac{2\pi}{\Lambda}\right)^2\Gamma^2(x)+\left(\frac{\partial\Gamma}{\partial x}\right)^2\right]\bm{\psi}, \label{eq:averaged}
\end{eqnarray}
in which we defined the refractive index matrix $\bm{N}=\sqrt{\bm{\epsilon}_D}=(n_o, 0; 0, n_e)$. We now divide all terms in Eq.\ (\ref{eq:averaged}) by $2\bar{n}k_0$ and take the two formal limits $\lambda/z_R \rightarrow 0$ (SVEA) and $\lambda/\Lambda = \Delta n \rightarrow 0$ (small birefringence approximation), while keeping $\Lambda$ and $z_R$ to finite values (this step and the subsequent analyses of the relative magnitude of various terms are best done after switching to dimensionless coordinates $x/w$ and $z/z_R$, but for the sake of brevity we keep here the dimensional ones). We thus obtain
\begin{eqnarray}
i\frac{\partial\bm{\psi}}{\partial z} && =
-\frac{1}{2\bar{n}k_0}\frac{\partial^2\bm{\psi}}{\partial x^2}-\frac{\pi}{\Lambda}\Gamma(x)\bm{S}_2\cdot\bm{\psi} +\frac{i}{2\bar{n}k_0}\frac{\partial\Gamma}{\partial x}\bm{S}_1\cdot\frac{\partial\bm{\psi}}{\partial x} +\frac{i}{4\bar{n}k_0}\left[\frac{\partial^2\Gamma}{\partial x^2}-\left(\frac{2\pi}{\Lambda}\right)^2\Gamma(x)\right]\bm{S}_1\cdot\bm{\psi}\nonumber\\
&& +\frac{1}{4\bar{n}k_0}\left[\left(\frac{2\pi}{\Lambda}\right)^2\Gamma^2(x)+\left(\frac{\partial\Gamma}{\partial x}\right)^2\right]\bm{\psi}.
\end{eqnarray}
Analyzing the magnitude of the four terms containing the optic axis perturbation $\Gamma(x)$, we can see that the last three are of order $\Lambda/z_R$ relative to the first. Hence, as a zero-order approximation in $\Lambda/z_R$, we can drop the last three terms, obtaining
\begin{equation}
i\frac{\partial\bm{\psi}}{\partial z} = -\frac{1}{2\bar{n}k_0}\frac{\partial^2\bm{\psi}}{\partial x^2}-\frac{\pi}{\Lambda}\Gamma(x)\bm{S}_2\cdot\bm{\psi}.
\end{equation}
In this equation the only nondiagonal matrix is $\bm{S}_2$. Its eigenvectors are obviously the circular polarizations $\bm{\psi}_R$ and $\bm{\psi}_L$ (with reference to the input plane), with eigenvalues $s_z=\pm1$, corresponding to the photon spin along $z$ (in $\hbar$ units). Hence, setting $\bm{\psi}(x,z)= \bm{\psi}_P A(x,z)$, with $P=L$ or $P=R$ for the two circular-polarized input waves, we obtain the following final amplitude propagation equation:
\begin{equation}
i\frac{\partial A}{\partial z} = -\frac{1}{2\bar{n}k_0}\frac{\partial^2 A}{\partial x^2}-s_z \frac{\pi}{\Lambda}\Gamma(x) A,
\end{equation}
where $s_z=+1$ for input RCP and $s_z=-1$ for input LCP, respectively.

We can now reconsider perturbatively the contribution of the three omitted terms in $\Gamma(x)$ which we had previously dropped. The second and third terms, linear in $\Gamma(x)$, include the matrix $\bm{S}_1$ which flips the CP handedness, so they are off-diagonal in the CP basis. The term quadratic in $\Gamma(x)$ is instead scalar and therefore diagonal in the CP basis (and in any other basis). Hence, the latter is the only relevant contribution to first order in $\Lambda/z_R$, while the other two contribute to the eigenvalues only to order $(\Lambda/z_R)^2$. In conclusion, up to first order in $\Lambda/z_R$, we can consider the following final expression for the photonic potential
\begin{equation}
V(x) = -s_z \frac{\pi}{\Lambda}\Gamma(x) + \frac{1}{4\bar{n}k_0}\left[\left(\frac{2\pi}{\Lambda}\right)^2\Gamma^2(x)+\left(\frac{\partial\Gamma}{\partial x}\right)^2\right].
\label{eq:fullphotonicpotential}
\end{equation}
The guided modes of the system can then be obtained by setting $A(x,z)=e^{i\beta z}A(x,0)$, with the effective propagation constant $\beta$ acting as an eigenvalue. The complete vector expression of the field in the $oe$ basis is given by $\bm{\psi}_{oe}(x,z)=\bm{U}_{oe}(z)\cdot \bm{\psi}_P A(x,z)$ with $P=L$ or $P=R$. To obtain the fields in the fixed $xy$ basis one needs to apply also the rotation matrix $\bm{R}_{xy}^{-1}(x,z)$. It should be noted that, when applied to a single input CP guided mode, this predicted evolution will keep the relative amplitudes of the ordinary and extraordinary waves constantly balanced everywhere in the medium. This justifies our statement that the average refractive index is unperturbed, so that only geometric phases affect the wave confinement.

In the Supplementary Material, a more formal theory based on the Bloch-Floquet method is adopted in order to analyze additional effects we might have neglected in the $z$-averaging operation. To first order in $\Lambda/z_R$, this more complete analysis returns the same results reported here. The quadratic contribution in $\Gamma$ appearing in Eq.\ (\ref{eq:fullphotonicpotential}) can be neglected for maximum $\Gamma$ values, that is $\Gamma_0$, up to 360$^\circ$ when the transverse size $w \approx 5$ $\mu$m, as apparent in Fig.\ \ref{fig:simulations}a-b. This term, however, becomes relevant for narrower distributions of $\theta$ (Supplementary Fig.\ 1 shows how the trapping potential gets strongly distorted for $\Gamma_0=360^\circ$ when $w_D=0.5$ $\mu$m). Finally, the possible long-term role of the higher-order contributions in $\Lambda/z_R$, neglected here, will be investigated in future work.

\subsection*{FDTD simulations}
For the FDTD numerical simulations we employed the open-source code MEEP \cite{OskooiRo10} to solve the full Maxwell equations in two dimensions, with no approximations. In all simulations we assumed a Gaussian shaped orientation of the optic axis across $x$ in the form $\Gamma(x)=\Gamma_0 \exp{(-x^2/w_D^2)}$. 
The excitation was a continuous-wave source launched in $x=z=0$ with a width  of 3 $\mu$m across $x$, turned on at $t=0$ and infinitely narrow across $z$. The modulated uniaxial medium was placed in $z_0=2$ $\mu$m with modulation $\sigma(z)=\sin\left[\frac{2\pi \Delta n}{\lambda}(z-z_0)\right]$. The refractive indices $n_o$ and $n_e$ were taken equal to 1.5 and 1.7, respectively. The simulations confirm that an input RCP is confined in the anisotropic structure, whereas an LCP input is expelled towards the edges (Supplementary Fig.\ 3). For the confined case, the polarization of the simulated wave undergoes small variations in the transverse plane (Supplementary Fig.\ 2) and is not perfectly periodic along $z$. Such small discrepancies between simulations and our analytic theory are clearly due to higher-order terms in $\lambda/\Lambda$ and $\Lambda/z_R$ which have been neglected in the latter. For $w_D=5$ $\mu$m light confinement improves with the maximum rotation $\Gamma_0$ (Supplementary Fig.\ 3); however, for $\Gamma_0=90^\circ$ even in the defocusing case a small amount of power is trapped on axis owing to the higher order terms appearing in Eq.\ (\ref{eq:fullphotonicpotential}). Further simulations show how light trapping/repulsion both increase as the width $w_D$ gets smaller, in agreement with theory (Supplementary Fig.\ 5). We also studied numerically the effects of shifting $\sigma(z)$ by changing $z_0$, obtaining a perfect agreement with theory. In fact, when $z_0=z_\mathrm{coh}/2$ the two polarization states exchange their roles, that is, LCP at the input gets trapped whereas a launched RCP undergoes defocusing (Supplementary Fig.\ 6); when $z_0=z_\mathrm{coh}/4$ the beam evolution does not depend on the input polarization, with power equally shared by confined and radiated modes. These results demonstrate that the PB-phase waveguide infringes translational symmetry along the propagation axis, at variance with standard (TIR or GRIN) waveguides. 
Finally, we numerically ascertained the role of small mismatches between the modulation period $\Lambda$ and the length $z_\mathrm{coh}$, an important issue in actual implementations of the proposed waveguides. Keeping  all the parameters fixed except for $\Lambda$, the global behaviour of the two polarization states, i.e., LCP defocused and RCP guided, is robust against $\Lambda$ variations up to 50$\%$. 

\subsection*{Fabrication of the geometric phase lenses}
Geometric phase lenses and similar PBOEs can be realized with a variety of techniques and materials  \cite{bomzon01a,marrucci06,slussarenko11,tabiryan09,alexeyev12}. Liquid crystals \cite{marrucci06} and liquid crystal polymers \cite{tabiryan09} are the most suitable for visible and near infrared illumination. Our GPLs were fabricated using polarization holography in combination with photo-alignment of nematic liquid crystal \cite{book_chigrinov_PA}. Planar cells were realized with two glass substrates held parallel at a separation of 6 $\mu$m, previously coated with a convenient photoaligning surfactant \cite{book_chigrinov_PA}. Such substrates were exposed to collimated ultraviolet light with an inhomogeneous distribution of the linear polarization state. The polarization pattern was realized by coaxial superposition of two beams (473~nm diode-pumped solid-state laser) with orthogonal circular polarizations and various phase-front curvatures. The interference of such beams, rather than an intensity modulation, produces a pattern of linear polarizations, with the angle of the polarization plane proportional to the point-wise phase difference between the beams. After exposure of the coated glass slides,  nematic liquid crystals (mixture E7 from Merck) were introduced in the cell by capillarity and aligned with a correspondingly inhomogeneous orientation of the optic axis (molecular director). The required half-wave phase retardation of the GPLs was finely adjusted by applying a 10 kHz square-wave electric voltage ($\approx$ 2.5 V peak-peak) as in other liquid-crystal-based PBOEs \cite{piccirillo10}. Each GPL can also be optically ``switched off'' by applying the voltage giving full-wave retardation ($\approx$ 4.0 V). Five GPLs were fabricated, with focal lengths $15.17$, $15.98$, $14.83$, $15.17$ and $15.69~\pm0.01$ cm, respectively, for light of wavelength 532 nm. The dispersion of the focal length values is due to the imperfect repeatability of the exposure conditions and insufficient stability of the fabrication setup.

\subsection*{Beam characterization within the structure}
In order to reconstruct the beam parameters inside the GPL structure, the propagating beam at 532 nm was sampled by a charge-coupled device (CCD) camera placed at various propagation distances between the lenses. Moreover, to improve the measurement accuracy of the local beam parameters, additional beam profiles were collected at given distances from each lens. This was accomplished by either switching off the GPLs following the one under measurement or by physically removing the remaining lenses from the sequence. The beam radii $w(z)$ were obtained through Gaussian fits of the acquired profile images and used to reconstruct the  modal parameters within and at the output of the waveguide. The obtained radius evolution was then compared with the theoretical predictions from ABCD Gaussian  propagation. A more realistic description of the beam was obtained with a non-unitary beam-quality factor $M^2$ of the confined beams after each lens ($M_1^2=1.05\pm0.01$,  $M_2^2=1.18\pm0.02$, $M_3^2=1.19\pm0.01$, $M_4^2=1.15\pm0.04$, and $M_5^2=1.32\pm0.04$, uncertainties at 95\% confidence level), suitably modifying the propagation equations \cite{sun98} for the simulations. The gradual increase of the $M^2$ parameter after each step of the discrete structure can be ascribed to degradation of the beam profile (as visible in Fig.\ \ref{fig:experiment}e) due to the noisy patterns of the GPLs. 


\newpage

\noindent\textbf{Acknowledgments}

\noindent The work in Naples was supported by the 7$^{th}$ Framework Programme of the European Union, within the Future Emerging Technologies program, under grant No.\ 255914, PHORBITECH, and within Horizon 2020 by the European Research Council (ERC), under ERC-advanced grant No.\ 694683, PHOSPhOR. A.A. and G.A. thank the Academy of Finland for financial support through the FiDiPro grant no. 282858. C.P.J. gratefully acknowledges Funda\c{c}\~{a}o para a Ci\^{e}ncia e a Tecnologia, POPH-QREN and FSE (FCT, Portugal) for the fellowship SFRH/BPD/77524/2011.

\vspace{1 EM}

\noindent\textbf{Author Contributions}

\noindent This work was jointly conceived by A.A., C.P.J., G.A. and L.M.; S.S. designed and carried out the experiment, with the help and supervision of B.P., E.S. and L.M.; A.A. and C.P.J. developed the theory and performed the numerical simulations, with the help and supervision of L.M. and G.A.; all authors discussed the results and contributed to the manuscript.

\vspace{1 EM}

\noindent\textbf{Additional Information}

\noindent Supplementary Information is included in the following pages. Correspondence and requests for materials should be addressed to L.M. or G.A.
\vspace{1 EM}

\noindent\textbf{Competing financial interests}

\noindent The authors declare no competing financial interests.




\clearpage

\onecolumngrid
\begin{center}
\noindent\textbf{SUPPLEMENTARY INFORMATION}
\end{center}
\renewcommand{\theequation}{S\arabic{equation}}
\renewcommand{\thesection}{S\arabic{section}}
\renewcommand{\thefigure}{S\arabic{figure}}
\renewcommand{\thepage}{S\arabic{page}}  
\setcounter{equation}{0}
\setcounter{section}{0}
\setcounter{figure}{0}
\setcounter{page}{1}
\vspace{1 EM}

We derive the model equations introduced in the main text and present numerical (FDTD) simulations in support of the most relevant results.
We also discuss a few additional features.

\section{Light propagation in a periodic system encompassing a rotation of the optic axis in the transverse plane}
\label{sec:homo_case}
We consider light propagation in inhomogeneous anisotropic dielectrics, in particular uniaxials; nonetheless, our results can be readily generalized to biaxial crystals. We take a medium whose dielectric properties vary across the transverse coordinate $x$ owing exclusively to a rotation of the optic axis in the plane $xy$ orthogonal to the propagation coordinate $z$. We consider finite wavepackets with wavevector parallel to $\hat{z}$; hence, corresponding to electric fields oscillating orthogonally and parallel to the optic axis, respectively, the two independent eigenvalues  $\epsilon_\bot$ and $\epsilon_\|$ of the relative permittivity tensor are constant in space. We  define the birefringence $\Delta n=\sqrt{\epsilon_\|}-\sqrt{\epsilon_\bot}=n_e-n_o$ and  assume that the distribution of the optic axis is purely planar and transverse to $\hat{z}$, such that  at each $z$ the rotation can be described by a standard 2D operator acting in $xy$:
\begin{equation}
  {R}_{xy}(\theta) = \left ( \begin{array} {cc}
  \cos\theta & \sin\theta \\
  -\sin\theta & \cos\theta \\
 \end{array} \right),
\end{equation}
where the angle $\theta$ is defined with respect to the $y$ axis of a Cartesian reference system in the laboratory frame, with $\theta=0$ corresponding to a dielectric permittivity $\epsilon_\|$ for electric fields along $y$. We study forward propagating light waves in the presence of a periodic modulation of $\theta$ along $z$. To this extent we set  
\begin{equation}
    \theta(x,z)= \sigma(z)\Gamma(x)=\left(  \sum_{p=-\infty}^\infty \sigma_p e^{\frac{i2\pi p z}{\Lambda}}  \right) \Gamma(x),
\end{equation}
with $\sigma(z)$ a function periodic with $\Lambda$ and $\Gamma(x)$ the transverse distribution of optic axis orientation, the latter being uniform across $y$. Hereafter, we focus on the case of a purely sinusoidal modulation $\sigma_p=\sigma_1 \delta_{1,p} + \sigma_{-1} \delta_{-1,p}$, with  $\delta_{p,p'}$ the Kronecker's delta and  $\sigma_{-1}=\sigma_1^*$ in order for $\sigma$ to be real valued. For the sake of simplicity, we refer to positive uniaxial media with $\Delta n>0$: the generalization to  negative birefringence is straightforward. Moreover, we deal with the case $\Lambda=\lambda/\Delta n$ in which resonant effects are expected to occur.

In the paraxial limit (neglecting longitudinal field components) and for small birefringence ($\Delta n \ll 1$),  Maxwell equations in two dimensions (i.e., no field evolution across $y$) can be cast as
\begin{equation} \label{eq:maxwell_equationS}
 \frac{\partial^2}{\partial z^2} \left ( \begin{array} {c}
  E_x \\
  E_y \\
 \end{array} \right)  = - \frac{\partial^2}{\partial x^2} \left ( \begin{array} {c}
  E_x \\
  E_y \\
\end{array} \right) - k_0^2 \left ( \begin{array} {cc}
  \epsilon_{xx}(x,z) & \epsilon_{xy}(x,z) \\
  \epsilon_{yx}(x,z) & \epsilon_{yy}(x,z) \\
 \end{array} \right) \left ( \begin{array} {c}
  E_x \\
  E_y \\
 \end{array} \right), 
\end{equation}
where $\bm{\uppsi}_{xy}=(E_x; E_y)$ is the two-component Jones vector representing the electric field in the complex notation.

In order to investigate the propagation of an electromagnetic (optical) wave in such a system, we make use of the transformation $\bm{\uppsi}_{oe}={R}_{xy}\cdot\bm{\uppsi}_{xy}$ and write $\bm{\uppsi}_{oe}=(E_o; E_e)$, with $E_o$ and $E_e$ the pointwise ordinary and extraordinary polarization components of the electric field, respectively. In the rotated reference system the two-dimensional dielectric tensor is diagonal, specifically ${\epsilon}_D=(\epsilon_\bot , 0 ; 0, \epsilon_\|)$. Eq.\ \eqref{eq:maxwell_equationS} then becomes
\begin{eqnarray}
 &&  \frac{\partial^2 \bm{\uppsi}_{oe}}{\partial z^2} - 2i \frac{\partial\theta}{\partial z} {S}_2 \cdot \frac{\partial \bm{\uppsi}_{oe}}{\partial z}   =
 -  \frac{\partial^2 \bm{\uppsi}_{oe}}{\partial x^2} + 2i  \frac{\partial\theta}{\partial x} {S}_2 \cdot \frac{\partial \bm{\uppsi}_{oe}}{\partial  x} +i \left( \frac{\partial^2 \theta}{\partial x^2} + \frac{\partial^2\theta}{\partial z^2} \right) {S}_2\cdot \bm{\uppsi}_{oe}   \nonumber \\
&& + \left[ \left(\frac{\partial\theta}{\partial x} \right)^2 + \left(\frac{\partial\theta}{\partial z}\right)^2  \right] \bm{\uppsi}_{oe} - k_0^2 {\epsilon}_D \cdot \bm{\uppsi}_{oe}, 
\label{eq:maxwell_rotated_inhoS}
\end{eqnarray}
where we introduced the Pauli matrix ${S}_2=\left(0,-i;i,0 \right)$. Equation \eqref{eq:maxwell_rotated_inhoS} shows that a scalar potential proportional to $(\partial\theta/\partial x)^2+(\partial\theta/\partial z)^2$ acts on both field components. Additionally, other terms  (containing the matrix ${S}_2$) couple ordinary and extraordinary polarizations: due to the rotation of the optic axis, the ordinary and extraordinary components are no longer independent, but can affect each other during propagation.

In the slowly varying envelope approximation (SVEA), we first set $E_o=\psi_o(x,z) e^{ik_0 n_o z}$ and $E_e=\psi_e(x,z) e^{ik_0 n_e z}$ in Eq.~\eqref{eq:maxwell_rotated_inhoS}. Exploiting the system periodicity in $z$ we then introduce the following further transformation:
\begin{equation}  \label{eq:ansatz_mode1}
  \psi_j (x,z) = A_j(x,z) B_j(x,z)  \ \ \ (j=e,o),
\end{equation}
where $A_j(x,z)$ are the slow-varying envelopes for the two wave components and $B_j(x,z)$ are periodic functions of $z$ accounting for the effect of the medium modulations with period $\Lambda$. It is convenient to write the latter ones in an exponential form as follows: 
\begin{equation}  \label{eq:ansatz_mode2}
  B_j(x,z) = \exp{{ \left[i {\int \sum_{\substack{p=-\infty \\ p\neq 0}}^\infty \beta_{p}^{(j)}(x,z)dz}\right]} } = \exp{ \left[\sum_{\substack{p=-\infty \\ p\neq 0}}^\infty  \frac{\Lambda}{2\pi p}\beta_p^{(j)}(x) e^{\frac{i2\pi pz}{\Lambda}} \right]}  
\end{equation}
where we set $\beta_p^{(j)}(x,z)=\beta_p^{(j)}(x)e^{\frac{i2\pi pz}{\Lambda}}$ and the constant term $p=0$ must be excluded from the sum. Inserting these ans\"atze in Eq.\ (\ref{eq:maxwell_rotated_inhoS}) and expanding in powers of $\Lambda$ all terms, one obtains a hierarchy of coupled equations for the amplitudes $A_j(x,z)$ and the coefficients $\beta_p^{(j)}(x)$. Neglecting from these equations all terms with powers of the period $\Lambda$ equal to or larger than 1 and higher-order terms in $\lambda/\Lambda$ (which is equal to $\Delta n$ at resonance),  we obtain the following coupled equations for the slow amplitudes:
\begin{align}
  i\frac{\partial A_o}{\partial z} = - \frac{1}{2\bar{n}k_0}\frac{\partial^2 A_o}{\partial x^2} + \frac{1}{2\bar{n}k_0} \left[ \overline{\sigma^2(z)}\left(\frac{d\Gamma}{dx} \right)^2 + \Gamma^2 \overline{\left(\frac{d\sigma}{dz}\right)^2}  \right] A_o + \frac{2\pi }{\Lambda} \sigma_{-1} \Gamma(x) A_e   , \label{eq:eigenvalue_E0_f1} \\
   i\frac{\partial A_e}{\partial z}  = - \frac{1}{2\bar{n}k_0}\frac{\partial^2 A_e}{\partial x^2} +  \frac{1}{2\bar{n}k_0} \left[ \overline{\sigma^2(z)} \left(\frac{d\Gamma}{dx} \right)^2 + \Gamma^2 \overline{\left(\frac{d\sigma}{dz}\right)^2}   \right] A_e  + \frac{2\pi}{\Lambda} \sigma_{1} \Gamma(x) A_0   ,  \label{eq:eigenvalue_E0_f2}
\end{align}
where $\overline{m(z)}=\frac{1}{\Lambda}\int_0^\Lambda m(z)dz$ and $\bar{n}=(n_e+n_o)/2$.

After inspection of Eqs. (\ref{eq:eigenvalue_E0_f1}-\ref{eq:eigenvalue_E0_f2}), three salient terms stand out:
\begin{itemize}
	\item a Kapitza-like potential, proportional to the squared transverse derivative of the distribution  $\Gamma(x)$ of optic axis orientation; its magnitude is modulated by the mean square of the periodic modulation $\sigma(z)$;    
	\item an effective transverse potential with profile proportional to $\Gamma^2(x)$; its magnitude is modulated by the mean square of the longitudinal derivative of the periodic modulation $\sigma(z)$;
	\item a phase-matched coupling between ordinary and extraordinary waves via the fundamental harmonics $\sigma_{\pm 1}$ of the periodic modulation.
\end{itemize}

The equations (\ref{eq:eigenvalue_E0_f1}-\ref{eq:eigenvalue_E0_f2}) can be recast in a more compact form as 
\begin{equation}  \label{eq:eigenvalue_vectorial}
  i \frac{\partial\textbf{A}}{\partial z} = {L}_\mathrm{ISO} \cdot \textbf{A} + {L}_\mathrm{ANI} \cdot \textbf{A},
\end{equation}
where $\textbf{A}=(A_o;\ A_e)$ and we introduced the isotropic (matrix) operator
\begin{equation}
 {L}_\mathrm{ISO} =\frac{1}{2\bar{n}k_0}\left\{ -\frac{\partial^2}{\partial x^2}  +  \left[  \overline{\sigma^2}\left(\frac{d\Gamma}{dx} \right)^2 +  \Gamma^2 \overline{\left(\frac{d\sigma}{dz}\right)^2} \right] \right\} {I}, \label{eq:def_LISO}
\end{equation} 
 and the anisotropic operator 
\begin{equation}
{L}_\mathrm{ANI} =  \frac{2\pi}{\Lambda} \Gamma(x) 
 \left( \begin{array}{cc}
   0 & \sigma_{-1} \\
    \sigma_{1} & 0  \\
 \end{array}  \right).
 \label{eq:def_LANI_complete}
\end{equation} 
The presence of ${L}_\mathrm{ANI}$ in Eq. \eqref{eq:eigenvalue_vectorial} accounts for the power exchange  between  ordinary and extraordinary components.

Let us now take $\sigma(z)=\sin(\frac{2\pi z}{\Lambda})$, that is $\sigma_1=-i/2$ and $\sigma_{-1}=i/2$; then ${L}_\mathrm{ANI}=-\frac{\pi}{\Lambda} \Gamma(x) {S}_2$. The eigenvalues of the Pauli matrix ${S}_2$ are $s_z = \pm 1$ with the two circular polarizations $(1;\ \pm i)/\sqrt{2}$ for eigenvectors (plus and minus correspond to RCP and LCP in our convention, respectively). Therefore, when the optic axis is modulated along $z$ with period equal to the beat length, the localized wave solutions are circularly polarized. By using the transformation
\begin{equation}
  \textbf{A}_{oe}={P}\cdot \textbf{A}_{LR} = \frac{1}{\sqrt{2}} 
  \left( \begin{array} {cc}
    1 & 1 \\
    -i & i
  \end{array} \right) \left( \begin{array} {c}
    A_L  \\
    A_R
  \end{array} \right),
\end{equation}
the system of equations separates into two independent scalar equations
\begin{align}  \label{eq:diagonal_LCP}
 i\frac{\partial A_L}{\partial z} = -\frac{1}{2\bar{n}k_0}\frac{\partial^2 A_L}{\partial x^2} + \frac{1}{4\bar{n}k_0}\left[\left(\frac{d\Gamma}{dx} \right)^2 + \frac{4\pi^2}{\Lambda^2} \Gamma^2(x)  \right] A_L + \frac{\pi }{\Lambda} \Gamma(x) A_L, \\
 i\frac{\partial A_R}{\partial z} = -\frac{1}{2\bar{n}k_0}\frac{\partial^2 A_R}{\partial x^2} + \frac{1}{4\bar{n}k_0}\left[\left(\frac{d\Gamma}{dx} \right)^2 + \frac{4\pi^2}{\Lambda^2} \Gamma^2(x)  \right] A_R - \frac{\pi }{\Lambda} \Gamma(x) A_R,
\end{align}
where we used $\overline{\sigma^2(z)}=0.5$ and $\overline{(d\sigma/dz)^2}=2\pi^2/\Lambda^2$.
The guided eigenmodes are then defined by setting $A_i(x,z)=e^{i\beta_0 z}A_i(x,0)$, where $\beta_0$ is the propagation constant. The problem is then transformed into a standard eigenvalue problem, with the following equations:
\begin{align}  \label{eq:eigenvalue_diagonal_LCP}
   -2k_0 \bar{n} \beta_0 A_L = - \frac{\partial^2 A_L}{\partial x^2} + \left[ \frac{1}{2}  \left(\frac{d\Gamma}{dx} \right)^2 + \frac{2\pi^2}{\Lambda^2} \Gamma^2(x)  \right] A_L + \frac{2k_0 \bar{n} \pi }{\Lambda} \Gamma(x) A_L,\\
   -2k_0 \bar{n} \beta_0 A_R = - \frac{\partial^2 A_R}{\partial x^2} + \left[ \frac{1}{2}  \left(\frac{d\Gamma}{dx} \right)^2 + \frac{2\pi^2}{\Lambda^2} \Gamma^2(x) \right] A_R - \frac{2k_0 \bar{n} \pi }{\Lambda} \Gamma(x) A_R, \label{eq:eigenvalue_diagonal_RCP}
\end{align}
Equations~\eqref{eq:eigenvalue_diagonal_LCP} and \eqref{eq:eigenvalue_diagonal_RCP} are valid for left (LCP) and right (RCP) circularly polarized wavepackets at the input interface, respectively. It is noteworthy that RCP and LCP interchange roles if the modulation $\sigma(z)$ is shifted by half a period, in agreement with the intuitive picture provided in the main text.

\subsection{Polarization-dependent effective index well}
In standard (1+1)D graded-index waveguides, a generic transverse-electric mode $u$ satisfies $-2k_0 \bar{n} \kappa u = - \frac{\partial^2 u}{\partial x^2} - k_0^2 \Delta n^2 u$, with $\kappa$ the propagation constant. Thus, from Eqs. (\ref{eq:eigenvalue_diagonal_LCP}-\ref{eq:eigenvalue_diagonal_RCP}) it is apparent that the two circular polarizations perceive the effective photonic potential $V_{\mathrm{eff}}=-k_0\Delta n^2_\mathrm{eff}/(2\bar{n})$ with the index distributions $\Delta n^2_{\mathrm{eff}}(x)$ given by
\begin{align}  \label{eq:well_LCP}
	\Delta n^2_{\mathrm{eff},LCP} = -\frac{2\bar{n}}{k_0} V_{\mathrm{eff},LCP} = -\frac{1}{k_0^2} \left[ \frac{1}{2}  \left(\frac{d\Gamma}{dx} \right)^2 + \frac{2\pi^2}{\Lambda^2} \Gamma^2(x)  \right] + \bar{n}\frac{\lambda }{\Lambda} \Gamma(x) ,\\
  	\Delta n^2_{\mathrm{eff},RCP} = -\frac{2\bar{n}}{k_0} V_{\mathrm{eff},RCP}  = -\frac{1}{k_0^2}\ \left[ \frac{1}{2}  \left(\frac{d\Gamma}{dx} \right)^2 + \frac{2\pi^2}{\Lambda^2} \Gamma^2(x) \right] - \bar{n} \frac{\lambda }{\Lambda} \Gamma(x) , \label{eq:well_RCP}
\end{align}
The polarization independent term between square brackets in Eqs.~(\ref{eq:well_LCP}-\ref{eq:well_RCP}) is a Kapitza-like equivalent photonic potential stemming from transverse and longitudinal modulation of the rotating optic axis. Since this photonic potential is independent from $k_0=2\pi/\lambda$, by itself it would support guided modes with wavelength-independent profile [NOTE: The propagation constant remains wavelength dependent through the vacuum wavevector on the LHS of Eqs.~(\ref{eq:eigenvalue_diagonal_LCP}-\ref{eq:eigenvalue_diagonal_RCP})].

The last terms on the RHS of Eqs. (\ref{eq:well_LCP}-\ref{eq:well_RCP}), with opposite signs as wave handedness reverses, are responsible for the strong dependence of light evolution on input polarization: the periodic rotation of the optic axis along $z$ allows for an \textit{accumulation} of the Berry phase during propagation, leading to the appearance of a potential proportional to $\Gamma$. Moreover, since phase-matching requires $\Lambda=\lambda/\Delta n$, such potential is directly proportional to the medium birefringence. A quantitative study on the relevance of each term in Eqs.\ (\ref{eq:well_LCP}-\ref{eq:well_RCP}) is reported in the following section.

\subsection{Bell-shaped orientation distribution of the optic axis}
\label{sec:quantitative_comparison}
\begin{figure}[htbp]
\centering
\includegraphics[width=0.9\textwidth]{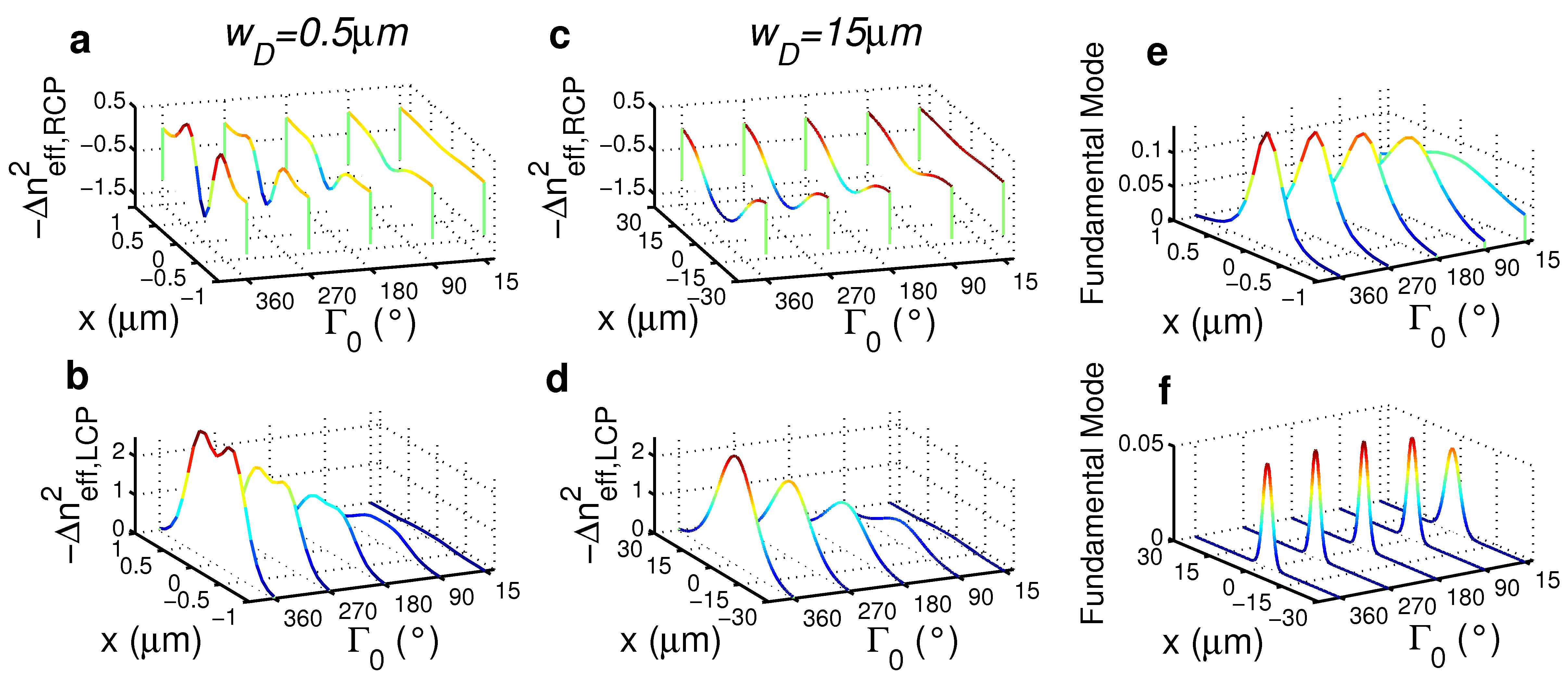}
\caption{ \textbf{a-d} Transverse profile of the sign-inverted effective index distribution, proportional to the photonic potential $V_{\mathrm{eff},LCP/RCP}$ versus maximum reorientation angle $\Gamma_0$ for two different widths (\textbf{a-b}) $w_D=0.5$ $\mu$m and  (\textbf{c-d}) $w_D=15$ $\mu$m of the transverse angular distribution. Input RCP waves are trapped in \textbf{a, c} and input LCP waves undergo defocusing in \textbf{b, d}, respectively, in agreement with the chosen initial section. \textbf{e-f} Profiles of the fundamental guided (RCP) mode when (\textbf{e}) $w_D=0.5$ $\mu$m  and (\textbf{f}) $w_D=15$ $\mu$m. Here $\lambda=1$ $\mu$m, $n_o=1.5$ and $n_e=1.7$.}
\label{fig:index_well_mode}
\end{figure}
Hereafter we make explicit reference to optical frequencies. Nevertheless, since Maxwell equations are invariant when dividing all the length scales by a given factor and multiplying the frequency by the same factor, our results clearly hold valid regardless of the electromagnetic band.

Supplementary Fig.\ \ref{fig:index_well_mode} shows the effective index well $-\Delta n^2_\mathrm{eff}$ (sign-inverted, so light is attracted by the dip) as computed from Eqs. (\ref{eq:well_LCP}-\ref{eq:well_RCP}) when the transverse distribution of the orientation angle $\Gamma(x)$ is bell-shaped and centered in $x=0$. We assumed $\Gamma(x)=\Gamma_0 \exp{(-x^2/w_D^2)}$, with $\Gamma_0$ and $w_D$ the maximum orientation angle and the width of the distribution, respectively. Using this simple ansatz we can address the role of each term in the  effective index well, Eqs. (\ref{eq:well_LCP}-\ref{eq:well_RCP}).  The term proportional to $(d\Gamma/dx)^2$ is a Kapitza-like term: when acting alone it yields quasi-modes, as detailed in Ref. \onlinecite{alberucci13}. This contribution to the index landscape, proportional to $w_D^{-2}$, increases as the angle distribution becomes narrower. Supplementary Fig.\ \ref{fig:index_well_mode} illustrates two  examples for $w_D=0.5$ $\mu$m  and $w_D=15$ $\mu$m, respectively: the Kapitza term becomes quite relevant for $w_D=0.5$ $\mu$m and large $\Gamma_0$, with the appearance of two local maxima, symmetrically placed with respect to the axis $x=0$.
The term proportional to $\Gamma^2(x)$ gives rise to anti-guidance, i.e. light repulsion from the symmetry axis. The term is proportional to the square of the birefringence $\Delta n$, thus dominates for large anisotropies.

Finally, the last term breaks the degeneracy between the two opposite circular polarizations. This term contributes with opposite signs to the overall potential acting on RCP and LCP waves, respectively: in the absence of other contributions, when the RCP (LCP) wave is confined, the LCP (RCP) is repelled from the region close to the symmetry axis $x=0$, i.e., it diffracts faster than in a homogeneous medium.

\subsection{Finite-Difference-Time-Domain numerical simulations of the guiding case}
In the numerical simulations we employed the open-source FDTD code named MEEP\cite{OskooiRo10} to solve the full Maxwell equations in two dimensions, assuming a Gaussian orientation of the optic axis across $x$, as above. The excitation was a continuous-wave source of wavelength 1 $\mu$m turned on at $t=0$, infinitely narrow across $z$ and launched in $x=z=0$. The source had a Gaussian profile of width 3 $\mu$m across $x$ and was point-like along $z$. The modulated uniaxial medium was placed in $z>z_0=2$ $\mu$m with modulation $\sigma(z)=\sin\left[\frac{2\pi \Delta n}{\lambda}(z-z_0)\right]$. The refractive indices $n_o$ and $n_e$ were taken equal to 1.5 and 1.7, respectively.

\begin{figure}[htbp]
\centering
\includegraphics[width=0.9\textwidth]{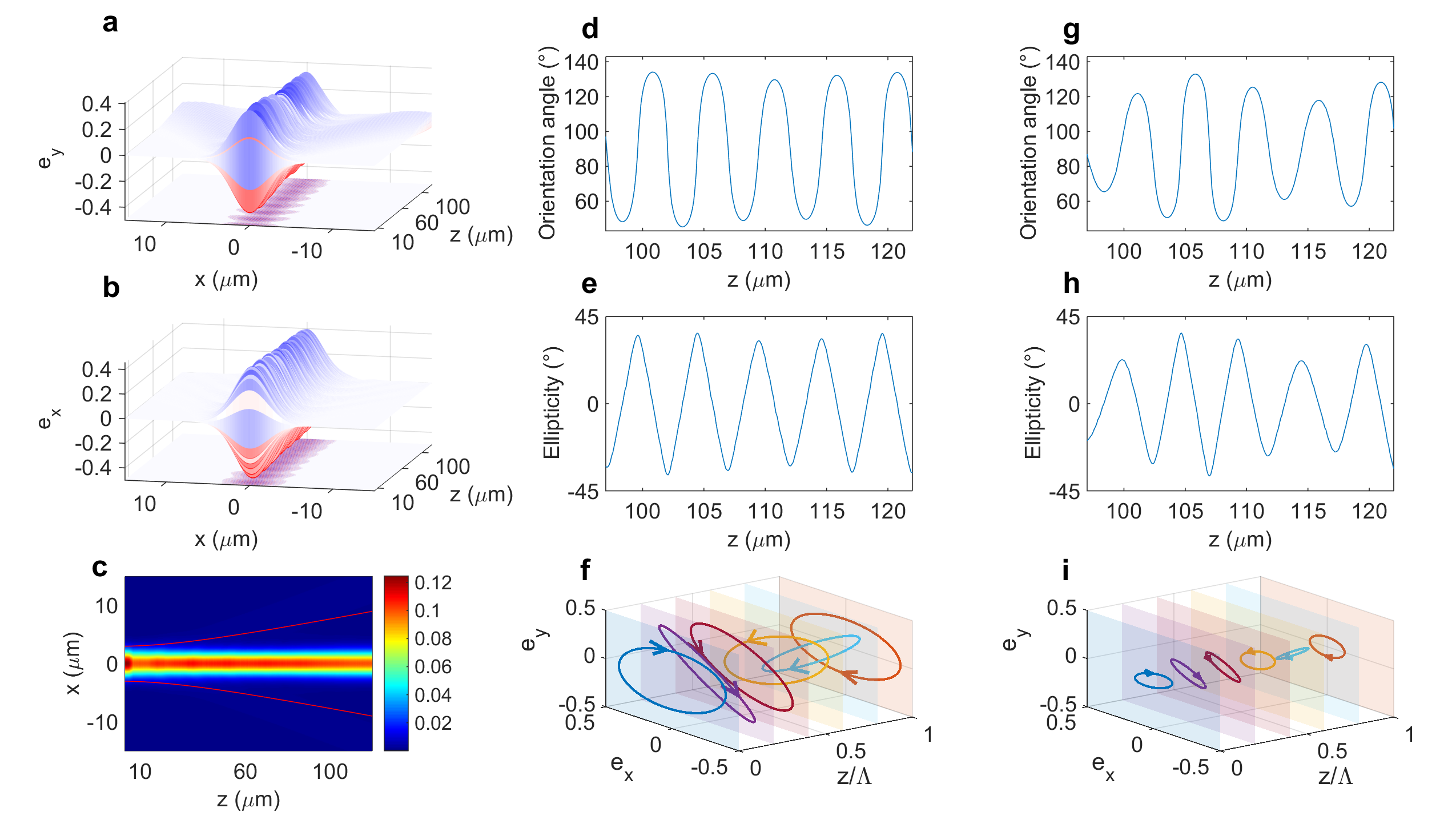}
\caption{Numerical snapshots of  (\textbf{a}) $e_y$ and (\textbf{b}) $e_x$  in the plane $xz$ once the stationary regime is reached in time;  (\textbf{c}) corresponding average intensity map in $xz$; the  red solid lines correspond to beam diffraction when $\Gamma_0=0$, i.e., the homogeneous case. (\textbf{d, g}) Angular orientation of the polarization ellipse and \textbf{e, h}) field ellipticity (arctangent of the ratio of minor to major axes) versus $z$ in (\textbf{d-e}) $x=0$ and (\textbf{g-h}) $x=3$ $\mu$m, respectively. Polarization ellipses in (\textbf{f}) $x=0$ and (\textbf{i}) $x=3$ $\mu$m  plotted versus $z$ in an interval $0 \leq z \Delta n/\lambda \leq 1$ far away from the excitation point, corresponding to 97 $\mu$m$\leq z\leq102$ $\mu$m. Here $w_D=5$ $\mu$m, $\Gamma_0=15^\circ$, RCP input. }
\label{fig:FDTD_sample}
\end{figure}

Supplementary Fig. \ref{fig:FDTD_sample} illustrates the propagation of an RCP wave input when $w_D=5$ $\mu$m and $\Gamma_0=15^\circ$, corresponding to a guiding potential (see Fig.\ 3 in the main text). As predicted, the natural diffractive spreading is compensated for by the effective waveguide resulting from Berry phase accumulation. At distances far enough from the input, in the stationary regime, the wavepacket acquires a periodic spatial distribution of its polarization and a nearly invariant profile. The polarization is generally elliptical; its trend can be examined by taking a single period $\lambda/\Delta n$ far enough from the input so that radiation (from imperfect coupling) is negligible. The polarization at the input is RCP, then it starts decreasing ellipticity; at about a quarter period it becomes linear, then elliptical again but with opposite handedness; at half period is nearly LCP, in excellent agreement with the theory. In the following half-period the polarization evolves in a similar manner, going from LCP to RCP. The polarization rotation versus propagation strictly resembles the behavior of a plane wave because it originates from the different phase velocities of the two linear eigenfield carriers (with $e^{ik_0 n_o z}$ and $e^{ik_0 n_e z}$, respectively). 

\begin{figure}[htbp]
\centering
\includegraphics[width=0.9\textwidth]{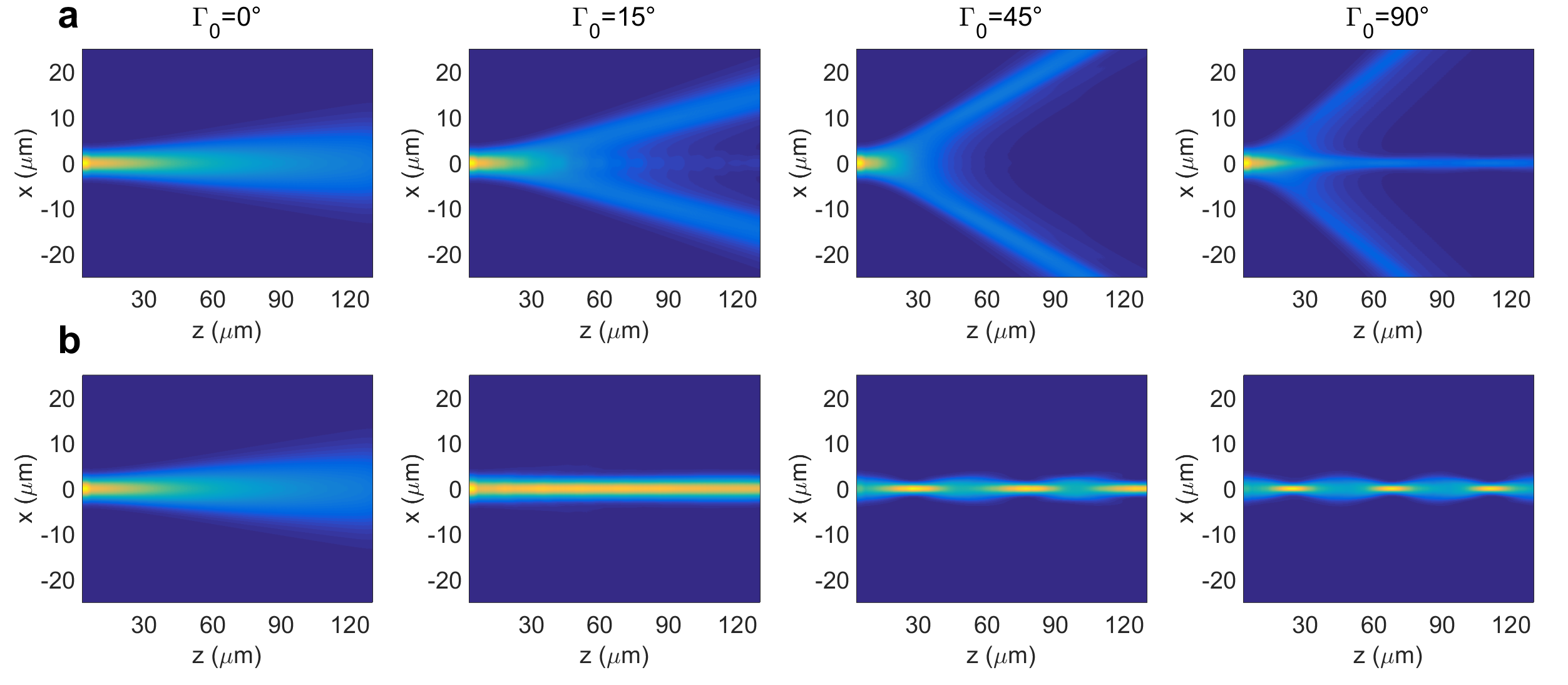}
\caption{Average beam intensity in the plane $xz$ when the input is (\textbf{a}) LCP  or (\textbf{b}) RCP. The maximum rotation $\Gamma_0$ is 0, 15, 45 and 90$^\circ$, from left to right, respectively. Here $z_0=0$ and $w_D=5$ $\mu$m.}
\label{fig:comparison_focusing_defocusing}
\end{figure}

Supplementary Fig.~\ref{fig:comparison_focusing_defocusing} compares the intensity evolution of propagating LCP and RCP wavepackets. In agreement with theory, one input polarization is subject to trapping, the other to a repulsive potential expelling light from the region around $x=0$. The strength of either potentials increases with the maximum rotation $\Gamma_0$: in the trapping case the waveguide eventually becomes multi-modal, as indicated by the appearance of breathing versus propagation; in the repulsive case the beam divergence increases with $\Gamma_0$. For $\Gamma_0 \geq 90^\circ$ the polarization of the confined beam continues to oscillate  along $z$, but in the presence of higher harmonics. In fact, in Eq. \eqref{eq:ansatz_mode2} the functions $\beta_p(x,z)$ with $p\geq 1$ must be accounted for and correspond to shorter periodicity in both beam profile and polarization. Simultaneously, a higher-order contribution proportional to $\Lambda^2 (\frac{d\Gamma}{dx}\frac{d^2 \Gamma}{dx^2})^2$ appears 
in the overall photonic potential of the isotropic operator ${L}_\mathrm{ISO}$ defined by Eq. \eqref{eq:def_LISO} [NOTE: This contribution is $O(\Lambda^2)$ and was neglected when deriving Eqs. (\ref{eq:eigenvalue_E0_f1}-\ref{eq:eigenvalue_E0_f2}): in the complete expression the term $\sum_{p>1} \left( \int{ \frac{\partial \beta^{(j)}_p}{\partial x} dz} \int{ \frac{\partial \beta^{(j)}_{-p}}{\partial x} dz} \right)\ (j=e,o)$ appears within square brackets and brings in the role of all the fast scales on the slow scale.].
Accordingly, even in the defocusing case a small portion of the wavepacket is trapped on-axis around $x=0$ for $\Gamma_0=90^\circ$, as visible in Supplementary Fig. \ref{fig:comparison_focusing_defocusing}. Owing to the additional modulating terms $\beta_p^{(j)}(x)$, further increases in $\Gamma_0$ (not shown) infringe the validity of Eqs. (\ref{eq:eigenvalue_diagonal_LCP}-\ref{eq:eigenvalue_diagonal_RCP}), leading to the generation of three-peaks for both input polarizations as well as appreciable changes of  polarization state across $x$ in the guided case.

\begin{figure}[htbp]
\centering
\includegraphics[width=0.9\textwidth]{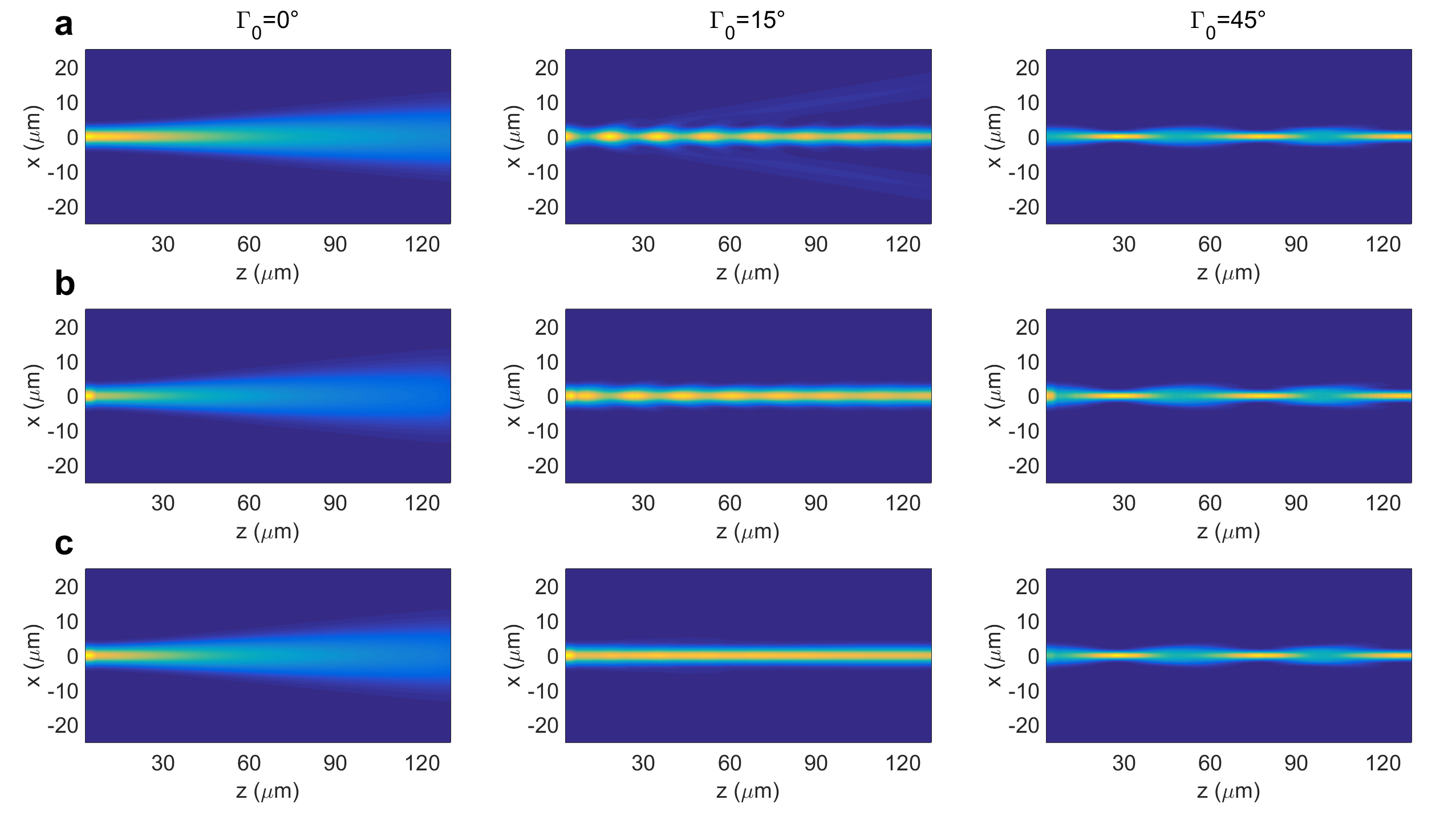}
\caption{Evolution maps of (\textbf{a}) $|E_y|^2$, (\textbf{b}) $|E_x|^2$ and (\textbf{c}) overall average intensity in the propagation plane $xz$. The maximum rotation $\Gamma_0$ is equal to 0, 15 and 45$^\circ$ from left to right, respectively. The input wave is RCP. Here $w_D=5$ $\mu$m.}
\label{fig:confined_case_vs_theta}
\end{figure}

Supplementary Fig.~\ref{fig:confined_case_vs_theta} shows how power is distributed between the two components $E_x$ and $E_y$ when light is guided:  $|E_x|^2$ and $|E_y|^2$ distributions in the plane $xz$  essentially coincide. The overall intensity is computed as $\bar{n}/(2Z_0)\left(|E_x|^2+|E_y|^2\right)$ with $Z_0$ the vacuum wave impedence, thus ignoring impedance variations for the two polarizations.

\subsection{Dependence on the transverse size of the effective waveguide}
\begin{figure}[htbp]
\centering
\includegraphics[width=0.9\textwidth]{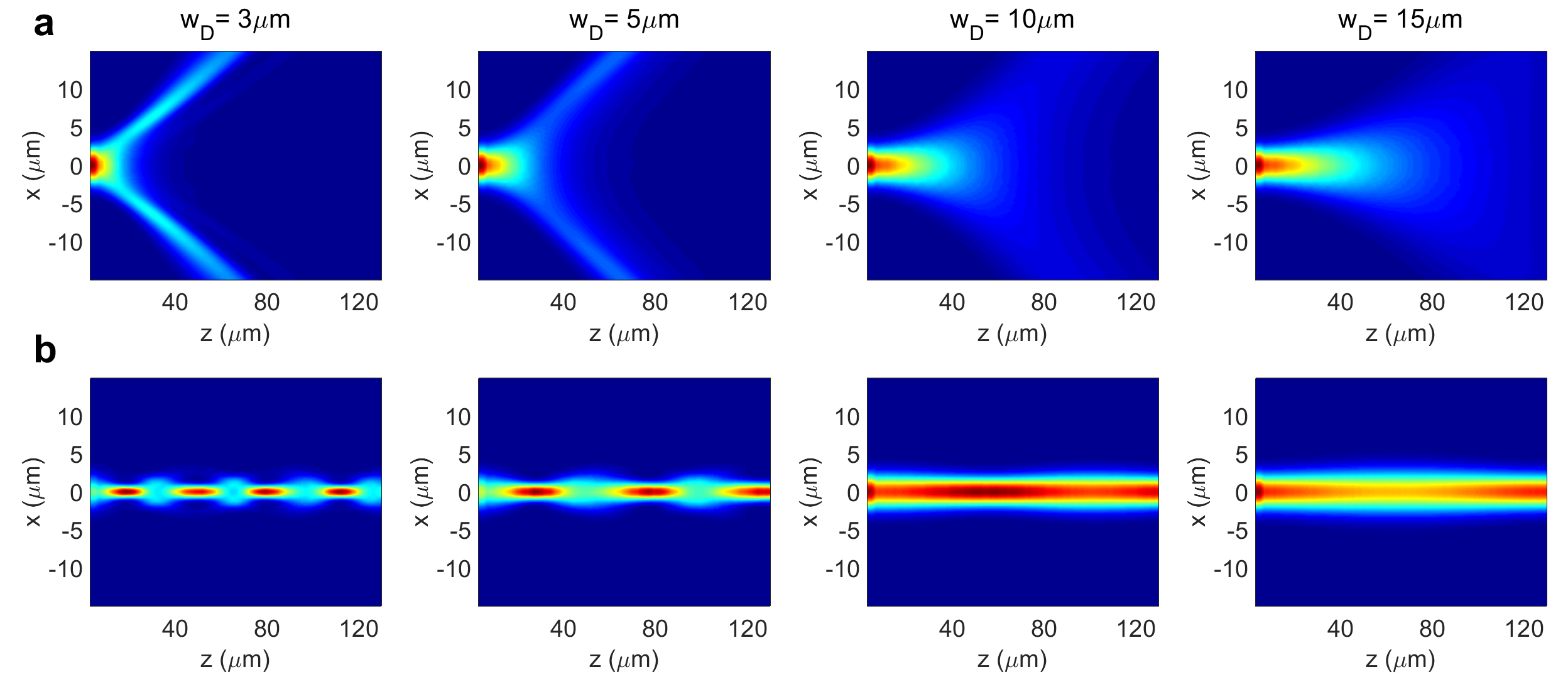}
\caption{Average intensity evolution in the propagation plane $xz$  for (\textbf{a}) left-  and (\textbf{b}) right-handed circularly polarized wavepackets. Here $\Gamma_0=45^\circ$ whereas the angular distribution width $w_D$ is 3, 5, 10 and 15 $\mu$m from left to right, respectively. }
\label{fig:confined_case_vs_width}
\end{figure}
 Another important feature of the system is its dependence on the transverse extent of the orientation angle distribution. Both circular polarizations are shown in Supplementary Fig.~\ref{fig:confined_case_vs_width}. In agreement with theory, the smaller $w_D$  the stronger is the repulsion of the defocused component from the perturbed region. Light spatial localization increases as $w_D$ reduces, with intensity oscillations becoming slower in space as the $\theta$ distribution  gets wider and wider. In line with Eqs. (\ref{eq:eigenvalue_diagonal_LCP}-\ref{eq:eigenvalue_diagonal_RCP}), light confinement undergoes the same trend as in standard waveguides based on total internal reflection.

\subsection{Dependence on the input point}
\begin{figure}[htbp]
\centering
\includegraphics[width=0.9\textwidth]{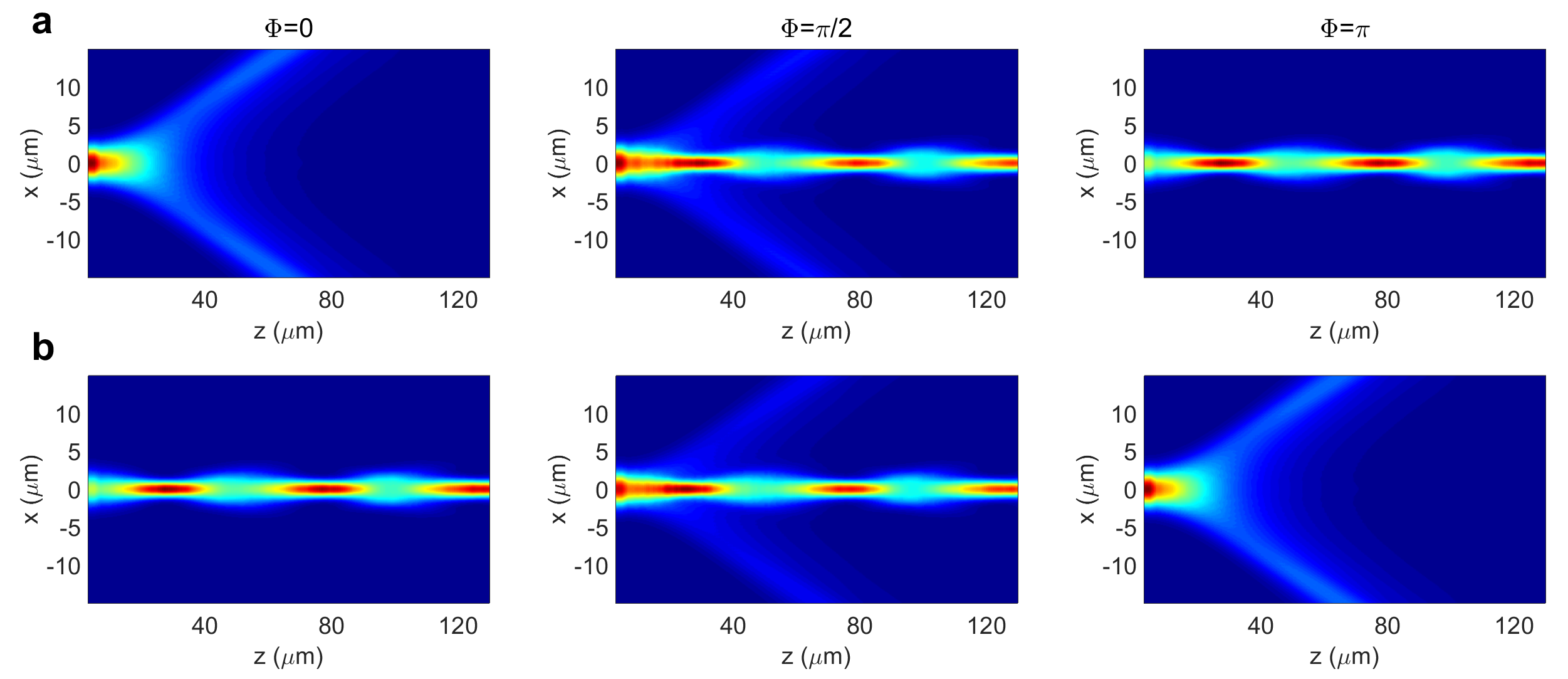}
\caption{ Wavepacket evolution in $xz$ for input LCP (\textbf{a}) and RCP (\textbf{b}) when the longitudinal modulation shift $\Phi$ is 0, $\pi/2$ and $\pi$, from left to right, respectively. Here $\Gamma_0=45^\circ$ and $w_D=5$ $\mu$m. }
\label{fig:evolution_vs_z0}
\end{figure}
Finally, we studied light propagation as the longitudinal modulation $\sigma(z)$ was shifted, that is, as the phase $\Phi$ in $\sigma=\sin\left[\frac{2\pi \Delta n}{\lambda}(z-z_0) + \Phi \right]$ was modified. Supplementary Fig. \ref{fig:evolution_vs_z0} shows the FDTD results: as predicted (Eqs. (\ref{eq:well_LCP}-\ref{eq:well_RCP})), when the phase is inverted (i.e., $\Phi=\pi$) the two circular polarizations exchange roles, with RCP waves going from trapping to anti-guiding and the opposite for LCP; when $\Phi=\pi/2$, the intensity evolution remains the same regardless of the input ellipticity: this agrees with Supplementary Fig. \ref{fig:FDTD_sample} showing quasi-linear polarization at a quarter period.
 
 
\end{document}